This is the ‘accepted manuscript’ version of our paper:

Main Controls on the Stable Carbon Isotope Composition of Speleothems

Jens Fohlmeister1,2,*, Ny Riavo G. Voarintsoa3, Franziska A. Lechleitner4, Meighan Boyd5, Susanne Brandtstätter6, Matthew J. Jacobson7, Jessica Oster8

1 Potsdam Institute for Climate Impact Research, Telegrafenberg, 14473 Potsdam, Germany

2 GFZ German Research Centre for Geosciences, Section 'Climate Dynamics and Landscape Development', Telegrafenberg, 14473 Potsdam, Germany

3 Department of Earth and Environmental Sciences, Katholieke Universiteit Leuven, Leuven, Belgium

4 Department of Earth Sciences, University of Oxford, South Parks Road, Oxford, OX1 3AN, UK

5 Department of Earth Sciences, Royal Holloway University of London, Egham, TW20 0EX, UK

6 Institute of Geology, University of Innsbruck, Innrain 52, 6020 Innsbruck, Austria

7 Department of Archaeology and Centre for Past Climate Change, University of Reading, Reading, RG6 6UR, UK

8 Department of Earth and Environmental Sciences, Vanderbilt University, Nashville, TN, 37240, USA.

* corresponding author; email: jens.fohlmeister@uni-potsdam.de

**Abstract**

The climatic controls on the stable carbon isotopic composition ($\delta^{13}C$) of speleothem carbonate are less often discussed in the scientific literature in contrast to the frequently used stable oxygen isotopes. Various local processes influence speleothem $\delta^{13}C$ values and confident and detailed interpretations of this proxy are often complex. A better understanding of speleothem $\delta^{13}C$ values is critical to improving the amount of information that can be gained from existing and future records. This contribution aims to disentangle the various processes governing speleothem $\delta^{13}C$ values and assess their relative importance. Using a large data set of previously published records we examine the spatial imprint of climate-related processes in speleothem $\delta^{13}C$ values deposited post-1900 CE, a period during which global temperature and climate data is readily available. Additionally, we investigate the causes for differences in average $\delta^{13}C$ values and growth rate under identical climatic conditions by analysing pairs of contemporaneously deposited speleothems from the same caves. This approach allows to focus on carbonate dissolution and fractionation processes during carbonate precipitation, which we evaluate using existing geochemical models. Our analysis of a large global data set of records reveals evidence for a temperature control, likely driven by vegetation and soil processes, on $\delta^{13}C$ values in recently deposited speleothems. Moreover, data-model intercomparison shows that calcite precipitation occurring along water flow paths prior to reaching the top of the speleothem can explain the wide $\delta^{13}C$ range observed for concurrently deposited samples from the same cave. We demonstrate that using the combined information of contemporaneously growing speleothems is a powerful tool to decipher controls on $\delta^{13}C$ values, which facilitates a more detailed discussion of speleothem $\delta^{13}C$ values as a proxy for climate conditions and local soil-karst processes.

## 1. Introduction

Speleothems, secondary cave carbonates, are valuable archives for reconstructing past climate conditions (e.g., Wong and Breecker, 2015). Of the various geochemical parameters they preserve, speleothem $\delta^{13}C$ values can provide information on climate and vegetation conditions (e.g., McDermott 2004; Ridley et al., 2015). However, it is often difficult to disentangle the various processes that influence speleothem $\delta^{13}C$ values, especially as some may be interdependent (e.g., vegetation type/density and rainfall amount/moisture availability). This complexity poses a significant challenge for the accurate interpretation of speleothem $\delta^{13}C$ values, and often limits its utility as a paleoclimate proxy to generalised discussions on overall conditions or broad vegetative transitions (e.g., Genty et al. 2003; Holmgren et al. 2003). However, in some cases $\delta^{13}C$ time series proved to be easier to interpret than corresponding $\delta^{18}O$ time series, in particular at sites where $\delta^{18}O$ is affected by multiple and competing effects (e.g., Genty et al., 2003; 2006; Scholz et al., 2012; Ridley et al., 2015; Mischel et al., 2017).

Although carbon transfer dynamics in cave systems have been extensively studied (e.g., Hendy, 1971; Genty et al., 1998; Oster et al., 2010; Fohlmeister et al., 2011; Rudzka et al., 2011; Lechleitner et al., 2016; Mattey et al., 2016; Bergel et al., 2017; Carlson et al., 2019), it is difficult to attribute individual processes as the driving force behind speleothem $\delta^{13}C$ variations (Griffiths et al., 2012; Spötl et al., 2016). The $\delta^{13}C$ composition of soil gas $CO_2$ is influenced by the type and density of vegetation above the cave, depending on the dominant photosynthetic pathway (C3, C4 or CAM plants), and by soil respiration rate (e.g., Cerling, 1984). The soil gas $CO_2$ is dissolved in percolating meteoric water, introducing a temperature-dependent fractionation effect on $\delta^{13}C$. This acidic water dissolves the underlying host rock carbonate until the solution is in equilibrium with respect to $Ca^{2+}$. The dissolution can occur under "open" conditions, where enough gaseous $CO_2$ is available in the soil or karst to allow for complete carbon exchange with the dissolved inorganic carbon species, or under "closed" conditions, where no gaseous $CO_2$ is present and carbon exchange is absent (Hendy, 1971). Intermediate conditions generally prevail in natural systems as often only a limited amount of gaseous $CO_2$ is available (e.g., Genty et al.,1998; Rudzka et al.,2011). Once the solution reaches the cave atmosphere with lower partial pressure of $CO_2$ ($pCO_2$), $CO_2$ starts to degas from the solution, triggering $CaCO_3$ precipitation. This process is accompanied by temperature dependent carbon isotope fractionation processes, which have been largely investigated by modelling studies (Dreybrodt et al., 2016; Dreybrodt & Scholz, 2011; Hansen et al., 2017; Mühlinghaus et al., 2009; Scholz et al., 2009) and laboratory studies (e.g., Polag et al., 2010; Wiedner et al., 2008; Hansen et al., 2019). Carbonate precipitation occurring before reaching the apex of the stalagmite is known as prior calcite precipitation (PCP) or, more rarely, prior aragonite precipitation. The degree of PCP usually depends on two parameters: I) on the $pCO_2$ gradient between the water and the gaseous phase and II) on length of the period the water in contact with the cave air before dripping.

There are several approaches to better constrain speleothem $\delta^{13}C$ variability. First, applying a multi-proxy approach that considers other proxies than $\delta^{13}C$, such as $\delta^{18}O$, radiocarbon or trace elements, and mineralogy or petrography, can shed light on the dominant processes that influence $\delta^{13}C$ variability in a given cave system (e.g., Oster et al., 2010; Rudzka et al., 2011; Griffiths et al., 2012; Fohlmeister et al., 2017; Voarintsoa et al., 2017c). Often, interpreting speleothem $\delta^{13}C$ values requires critical knowledge about the cave system, including potential anthropogenic impacts (e.g., Baldini et al., 2005; Mattey et al 2008; 2010; Hartmann et al., 2013; Burns et al., 2016; Voarintsoa et al., 2017c). A second approach lies in the investigation of the acting processes through analysis of a large and spatially extensive network of speleothem records (Breecker, 2016). This has the potential to remove highly localised, site-specific variability, and allow detection of more general relationships between speleothem $\delta^{13}C$ composition and climate or ecosystem conditions.

Here, we use speleothem $\delta^{13}C$ records compiled in the first version of the SISAL database (Atsawawaranunt et al., 2018a; b), henceforth denoted SISAL_v1. This database was compiled to provide a comprehensive understanding of speleothem $\delta^{18}O$ and $\delta^{13}C$ records for climate reconstruction and model evaluation (e.g., Comas-Bru et al., 2019). From this database, several papers have already been published that assess data coverage and investigate regional patterns in stalagmite $\delta^{18}O$ records from specific regions and continents (e.g., Lechleitner et al., 2018; Oster et al., 2019; Braun et al., 2019; Burstyn et al., 2019). To constrain the governing processes influencing speleothem $\delta^{13}C$ values in the SISAL records we focus on two subsets of the data of the SISAL_v1 database. First, we analyse globally distributed records that cover the period between 1900 and 2014 CE (hereafter denoted post-1900 CE), to investigate the spatial relationship between $\delta^{13}C$ and climate by comparing the speleothem records with available instrumental data. Second, we analyse records from contemporaneously growing speleothems from the same cave to shed light on karst and cave processes and to evaluate the utility of Rayleigh isotope fractionation models (e.g., Deininger et al., 2012; Deininger and Scholz, 2019) for understanding speleothem $\delta^{13}C$ records.

## 2. The SISAL_v1 data

### 2.1. Post- 1900 CE speleothem $\delta^{13}C$ data

We extracted $\delta^{13}C$ data of all speleothems that grew after 1900 CE from SISAL_v1, resulting in 59 speleothem records from 50 individual caves (Tab. 1) and yielding about 3600 individual $\delta^{13}C$ measurements. Three speleothems have only one data point in the according time window, but were nonetheless included in the analysis. The record with the highest number has 659 data points (stalagmite YOK-G, Ridley et al., 2015).

***Tab. 1:*** *Speleothem $\delta^{13}C$ records growing after 1900 CE extracted from SISAL_v1. Speleothems marked by an (*) consist of aragonite. Column 'Comments on dating': number of U-Th dating points in the post 1900 CE period / actively dripping or layer counting (LC) / evidence through radiocarbon bomb peak detection (14C-BPD) or $^{210}Pb$ dating.*

| Cave | Speleothem | number of data point | $\delta^{13}C$ [‰] | MAT [°C] | Precipi-tation [mm/a] | altitude [m asl] | Comments on dating | reference |
|---|---|---|---|---|---|---|---|---|
| Anjohibe Cave | AB3 | 12 | 3.38 | 27.2 | 1496 | 100 | 1/-/- | Burns et al., 2016 |
| Anjohibe Cave | AB2 | 191 | 3.49 | 27.2 | 1496 | 100 | 1/-/- | Scroxton et al., 20 |
| Ascunsa Cave | POM2 | 2 | -10.37 | 8.2 | 600 | 1050 | 0/dripping/- | Dragusin et al., 20 |
| Bero Cave | Bero-1 | 135 | -4.61 | 18.9 | 1030 | 1363 | 0/LC/14C-BPD | Asrat et al., 2008 |
| Bir-Uja Cave | Keklik1 | 58 | -4.89 | 12.1 | 353 | 1325 | 0/-/14C-BPD | Fohlmeister et al., |
| Botuverá Cave | BT-2 | 1 | -6.16 | 19.5 | 1300 | 230 | 0/-/- | Cruz et al., 2005 |
| Brown's Follymine | Boss | 17 | -9.63 | 10 | 842 | 150 | 0/LC/- | Baldini et al., 2005 |
| Brown's Follymine | BFM-9 | 12 | -9.31 | 10 | 842 | 150 | 0/LC/14C-BPD | Baldini et al., 2005 |
| Brown's Follymine | F2 | 17 | -8.83 | 10 | 842 | 150 | 0/LC/- | Baldini et al., 2005 |
| Bunker Cave | Bu4 | 8 | -5.86 | 10.8 | 950 | 184 | 0/-/14C-BPD | Fohlmeister et al., |
| Cango Cave | V3 | 2 | -6.34 | 17.5 | 172 | 650 | 0/dripping/- | Talma and Vogel, 1 |
| Ceremosnja Cave | CC-1 | 3 | -7.84 | 11.6 | 695 | 530 | 0/dripping/- | Kacinski et al., 200 |
| Cold Air Cave | T8* | 13 | -6.09 | 17.3 | 521 | 1420 | 0/-/- | Holmgren et al., 20 |
| Cold Air Cave | T7_2013* | 106 | -7.12 | 17.3 | 521 | 1420 | 1/LC/14C-BPD | Sundquist et al., 20 |
| Crag Cave | CC3 | 2 | -7.41 | 10.5 | 1475 | 60 | 0/-/- | McDermott et al., |
| Dante Cave | DP1_2016* | 31 | -10.70 | 21 | 532 | 1300 | 2/-/- | Voarintsoa et al., 2 |
| Defore Cave | S3 | 83 | -10.16 | 25.7 | 500 | 150 | 0/LC/- | Burns et al., 2002 |
| DeSoto Caverns | DSSG-4* | 33 | -9.50 | 17.4 | 1406 | 170 | 0/-/- | Aharon et al., 201 |
| Furong Cave | FR-0510* | 9 | -5.32 | 18.3 | 1086 | 260 | 0/-/- | Li et al., 2011 |
| Guillotine Cave | GT05-5 | 9 | -9.23 | 9.6 | 2400 | 740 | 0/-/- | Whittaker, 2008 |
| Han-sur-Lesse Cave | Han-stm5b | 15 | -9.37 | 8.9 | 787 | 180 | 0/LC/14C-BPD | Genty et al., 1998 |
| Heshang Cave | HS4_2008 | 102 | -12.08 | 18 | 1460 | 694 | 0/LC/14C-BPD | Hu et al., 2008 |
| Ifoulki Cave | IFK1 | 66 | -7.93 | 17 | 400 | 1265 | 1/-/- | Ait Brahim et al., 2 |
| Jhumar Cave | JHU-1 | 70 | -11.92 | 25.5 | 1503 | 600 | 1/-/- | Sinha et al., 2011 |
| Kesang Cave | KS08-1-H | 2 | -6.20 | 4.5 | 500 | 2000 | 0/-/- | Cheng et al., 2016 |
| Kinderlinskaya Cave | KC-3 | 2 | -8.16 | 3 | 560 | 240 | 0/-/- | Baker et al., 2017 |
| Klapferloch Cave | PFU6 | 38 | -2.54 | 4.8 | 600 | 1140 | 0/LC/- | Boch and Spötl, 20 |
| Korallgrottan Cave | K11 | 5 | -6.25 | 1.4 | 866 | 600 | 0/-/- | Sundqvist et al., 20 |
| Leviathan Cave | LC-1 | 3 | -4.18 | 8.3 | 106 | 2400 | 0/-/- | Lachniet et al., 201 |
| Liang Luar Cave | LR06-B1_2016 | 58 | -9.90 | 25 | 1200 | 550 | 1/-/14C-BPD | Griffiths et al., 201 |
| Macal Chasm | MC01 | 15 | -11.02 | 21 | 2095 | 550 | 0/active/210Pb | Webster et al., 200 |
| Modric Cave | MOD-22 | 6 | -7.25 | 16 | 960 | 32 | 0/dripping/- | Rudzka et al., 201 |
| Munagamanu Cave | Mun-stm2 | 44 | -5.72 | 27.6 | 526 | 475 | 1/-/- | Atsawawaranunt e |
| Munagamanu Cave | Mun-stm1 | 21 | -4.29 | 27.6 | 526 | 475 | 2/-/- | Atsawawaranunt e |
| Natural Bridge Caverns | NBJ | 6 | -7.71 | 21 | 740 | 315 | 0/-/- | Wong et al., 2015 |
| New St Michael's Cave | Gib04a | 443 | -11.04 | 18.3 | 767 | 426 | 0/LC/14C-BPD | Mattey et al., 200 |
| Okshola Cave | FM3 | 2 | -6.56 | 3.2 | 1000 | 165 | 0/-/- | Linge et al., 2009 |
| Palestina Cave | PAL3 | 10 | -11.40 | 22.8 | 1570 | 870 | 0/-/- | Apaéstegui et al., |
| Paraiso Cave | PAR03 | 19 | -8.97 | 26 | 2400 | 60 | 2/-/- | Wang et al., 2017 |
| Perdida Cave | PDR-1 | 86 | -5.53 | 27.3 | 1375 | 400 | 0/active/- | Winter et al., 201 |
| Postojna Cave | POS-STM-4 | 10 | -9.59 | 8 | 1500 | 529 | 0/event/14C-BPD | Genty et al., 1998 |
| Rukiessa Cave | Merc-1 | 84 | -5.15 | 18.9 | 1030 | 1618 | 0/LC/14C-BPD | Baker et al., 2007 |
| Rukiessa Cave | Asfa-3 | 83 | -6.44 | 18.9 | 1030 | 1618 | 0/LC/14C-BPD | Baker et al., 2007 |
| Sahiya Cave | SAH-AB | 110 | 0.24 | 22 | 1600 | 1190 | 1/-/- | Sinha et al., 2015 |
| Sofular Cave | So-1 | 56 | -9.57 | 13.8 | 1200 | 400 | 0/-/- | Fleitmann et al., 2 |
| Soreq Cave | Soreq-composite | 20 | -10.17 | 20 | 500 | 400 | 0/-/- | Grant et al., 2012 |
| Soylegrotta Cave | SG95 | 4 | -3.60 | 3.5 | 1450 | 280 | 0/-/- | Linget al., 2001 |
| Tamboril Cave | TM0* | 10 | -11.09 | 22.5 | 1400 | 575 | 1/-/- | Wortham et al., 20 |
| Taurius Cave | Taurius | 293 | -11.71 | 26 | 2735 | 230 | 4/-/210Pb | Partin et al., 2013 |
| Tonnelnaya Cave | TON-2 | 2 | 1.38 | 3.1 | 355 | 3226 | 0/-/- | Cheng et al., 2016 |
| Uamh an Tartair | SU967 | 39 | -11.16 | 7.1 | 1900 | 220 | -/LC/- | Baker et al., 2012 |
| Uamh an Tartair | SU032 | 101 | -11.89 | 7.1 | 1900 | 220 | -/LC/- | Baker et al., 2011 |
| Ursilor Cave | PU-2 | 1 | -10.81 | 9.7 | 950 | 482 | 0/-/- | Onac et al., 2002 |
| Villars Cave | Vil-stm6 | 1 | -9.06 | 12.5 | 1005 | 175 | 0/-/- | Atsawawaranunt e |
| Villars Cave | Vil-stm1 | 21 | -8.56 | 12.5 | 1005 | 175 | 0/LC/14C-BPD | Labuhn et al., 2015 |
| Wah Shikhar Cave | WS-B | 72 | -2.59 | 17 | 2150 | 1290 | 0/-/- | Sinha et al., 2011 |
| Xinya Cave | XY07-8 | 75 | -4.74 | 16.1 | 1130 | 1250 | 0/dripping/- | Li et al., 2017 |
| Yok Balum Cave | YOKI* | 174 | -9.86 | 22.9 | 2950 | 366 | 3/-/14C-BPD | Kennett et al., 201 |
| Yok Balum Cave | YOKG* | 659 | -10.52 | 22.9 | 2950 | 366 | 6/-/14C-BPD | Ridley et al., 2015 |

### 2.2. Speleothem $\delta^{13}C$ data from contemporaneous samples

The second data sub-set extracted from SISAL_v1 comprises speleothems which grew at the same time in the same cave for at least 20 consecutive years, without upper limitations on the time period of growth. In total, 94 speleothems from 32 caves fulfil this requirement (Tab. 2). Up to seven at least pairwise coevally growing speleothems from a single cave are available. With our search criteria we extracted approximately 57,000 individual $\delta^{13}C$ data points. For each data point we also extracted its depth along the speleothem profile and its age.

***Tab. 2:*** $\delta^{13}C$ *records from speleothems growing contemporaneously in the same cave from SISAL_v1. The records are grouped by cave, and the number of overlapping samples is indicated.*

| Cave | Stalagmites | Number of overlap (#) | Reference |
|---|---|---|---|
| Abaco Island Cave | AB-DC-01, AB-DC-03, AB-DC-09 | 2 | Arienzo et al., 2017 |
| Abaliget Cave | ABA_1, ABA_2 | 1 | Koltai et al., 2017 |
| Anjohibe Cave | AB3, AB2**, MA3**, ANJB-2** | 6 | Scroxton et al., 2017; Voarintsoa et al., 2017b,c; Burns et al., 2016 |
| Antro del Corchia | CC-1_2009, CC-5_2009, CC-7 | 2 | Drysdale et al., 2009 |
| Baradla Cave | BAR-IIL,BAR-IIB | 1 | Demény et al., 2017 |
| Bittoo Cave | BT-1, BT-2.1, BT-2.2 | 2 | Kathayat et al., 2016 |
| Botuverá Cave | BTV21a, BT-2 | 1 | Bernal et al., 2016; Cruz et al., 2005 |
| Brown's Follymine | Boss, F2, BFM-9 | 3 | Baldini et al., 2005 |
| Buckeye Creek | BCC-8, BCC-10 | 1 | Springer et al., 2014 |
| Bunker Cave | Bu1, Bu2, Bu4, Bu6 | 3 | Fohlmeister et al., 2012 |
| Devils Hole | DH2, DH2-D, DH2-ETerminal1, DH2-ETerminal2 | 4 | Moseley et al., 2016 |
| Dim Cave | Dim-E2, Dim-E3, Dim-E4 | 3 | Ünal-İmer et al., 2015 |
| Gueldaman Cave | stm2, stm4 | 1 | Ruan et al., 2016 |
| Katerloch Cave | K1, K3 | 1 | Boch et al., 2009 |
| Kesang Cave | KS06-A-H, KS08-2-H, KS06-A, KS06-B, KS08-1, KS08-2, | 5 | Cheng et al., 2016 |
| Kinderlinskaya Cave | KC-1, KC-3 | 1 | Baker et al., 2017 |
| Lancaster Hole | LH-70s-1, LH-70s-2, LH-70s-3 | 3 | Atkinson and Hopley, 2013; Atsawawaranunt et al., 2018a |
| Liang Luar Cave | LR07-A8, LR07-A9, LR07-E11, LR06-B1_2016, LR06-B3_2016 | 4 | Griffiths et al., 2013; 2016 |
| Mairs Cave | MC-S1, MC-S2 | 1 | Treble et al., 2017 |
| Milchbach Cave | MB-2, MB-3, MB-5 | 3 | Luetscher et al., 2011 |
| Molinos Cave | Mo-1, Mo-7 | 1 | Moreno et al., 2017 |
| Munagamanu Cave | Mun-stm1, Mun-stm2** | 1 | Atsawawaranunt et al., 2018a |
| Okshola Cave | FM3, Oks82 | 1 | Linge et al., 2009 |
| Palestina Cave | PAL3, PAL4 | 1 | Apaéstegui et al., 2014 |
| Paraiso Cave | PAR01, PAR03, PAR06, PAR07, PAR08, PAR16, PAR24 | 7 | Wang et al., 2017 |
| Rukiessa Cave | Merc-1, Asfa-3 | 1 | Baker et al., 2007 |
| Tamboril Cave | TM0*, TM2* | 1 | Worthham et al., 2017 |
| Tonnel'naya Cave | TON-1, TON-2 | 1 | Cheng et al., 2016 |
| Uamh an Tartair | SU967, SU032 | 1 | Baker et al., 2011; 2012 |
| Villars Cave | Vil-stm1, Vil-stm6, Vil-stm9, Vil-stm11, Vil-stm14, Vil-stm27, Vil-car1 | 11 | Labuhn et al., 2015; Wainer et al., 2011; Genty et al., 2003; 2006; 2013; Atsawawaranunt et al., 2018a |
| Yok Balum Cave | YOK-I*, YOK-G* | 1 | Kennett et al., 2012; Ridley et al., 2015 |
| Zhuliuping Cave | ZLP1, ZLP2 | 1 | Huang et al., 2016 |

*Footnotes: (*) All speleothems from the cave are entirely aragonitic. (**) Both aragonitic and calcitic stalagmites are available from this cave, or mineralogical transitions within a single sample are present. (#) Three stalagmites A, B, C might provide two (A-B, A-C but not B-C.) or three (A-B, A-C and B-C) periods of overlap. For four speleothems it is possible that only two periods of contemporaneous growth occur (two pairs of stalagmites with contemporaneous growth, but the first stalagmite pair itself does not have an overlap with the second pair) or up to six (all stalagmites grew at the same time).*

The shortest time period of overlap is about 100 years for Rukiessa Cave (Baker et al., 2007) and the longest period of overlap is for the Kesang Cave (Cheng et al., 2016), where stalagmites grew contemporaneously for about 50000 years. Most stalagmite pairs grew contemporaneously for 1000 to 10000 years. The number of $\delta^{13}C$ data points per stalagmite in coeval growth phases is between 7 (Paraiso Cave; Wang et al., 2017) and more than 3500 (Yok Balum Cave, Ridley et al., 2015). Most stalagmites provide data between 100 to 300 data points for the coeval growth phases.

## 3. Methods

### 3.1. Climate influence on speleothem $\delta^{13}C$

We analyse the SISAL_v1 $\delta^{13}C$ dataset covering the period post-1900 CE and compare it to instrumental climate data to investigate the influence of temperature, rainfall amount, altitude or vegetation on speleothem $\delta^{13}C$ values. We calculate the average $\delta^{13}C$ value and its variance, to prevent over-interpretation of individual $\delta^{13}C$ values that may contain temporally restricted outliers. As carbon isotope fractionation effects are different for aragonite and calcite, the $\delta^{13}C$ values of aragonitic speleothems (Tab. 2) were corrected to the corresponding $\delta^{13}C$ value of calcite by using the fractionation offset between both polymorphs established in a laboratory study (Romanek et al., 1992). This fractionation offset was recently confirmed by a speleothem study, where calcite-aragonite transitions occurred along individual growth layers (Fohlmeister et al., 2018).

Climate data, i.e., present-day mean annual temperature, rainfall amount, and altitude (Tab. 1), were obtained either directly from the original papers or - where necessary - from companion papers reporting on the same cave. Vegetation data is derived from SPOT-VEGETATION satellite imagery from the Global Land Cover 2000 Project (GLC2000), used in Aaron and Gibbs (2008). The data, available online from the Carbon Dioxide Information Analysis Center (http://cdias.ess-dive.lbl.gov), were imported to ArcGIS 10.5 as shapefile, and plotted using the coordinate reference system WGS-84. To ease comparison with speleothem $\delta^{13}C$ averages, we grouped each polygon based on vegetation cover categories (e.g., steppe, dry forest, shrub land, rainforest). The average speleothem $\delta^{13}C$ values were binned into 1‰ increments, yielding ten categories, ranging from data lower than -11.00 ‰ VPDB (Vienna PeeDee Belemnite) to data higher than -3.00 ‰ VPDB.

It is important to note that age uncertainties related to either U-Th dating uncertainties and/or interpolation techniques may complicate interpretation of the results from this study. To address this, however, we have tested how chronological uncertainty may influence $\delta^{13}C$ values, by calculating the d13C averages of the last 100 years (i.e., back to 1919AD), in comparison with the interval back to 1900 AD (our initial boundary set for this study). This 19-year change in the applied time interval refers to a change in relative age uncertainty of 16% and thus well represents the typical U-Th derived age uncertainty during this period. This simple test suggests that the average $\delta^{13}C$ values did not change by more than 0.1 ‰ between the two intervals studied. Therefore, we presume that chronological uncertainty is insignificant at this temporal scale.

### 3.2. Influence of local processes on speleothem $\delta^{13}C$ values

The investigation of karst and fractionation processes during $CaCO_3$ dissolution and precipitation is more technical. Contemporaneously growing speleothems from the same cave were used on the assumption that temperature, rainfall amount and vegetation above the cave should influence their $\delta^{13}C$ value in a similar way. In that case, any variability in $\delta^{13}C$ values between individual speleothems must be related to the conditions under which carbonate is dissolved and precipitated.

To explore this, we use the second dataset extracted from SISAL_v1, and calculate the average growth rate along with the average and standard deviation of $\delta^{13}C$ of contemporaneous growth sections in the speleothems. For the growth rate, the standard deviation could not be calculated as unfortunately most SISAL_v1 records lack information on interpolated age uncertainties. Where speleothems from individual caves exist in calcitic and aragonitic form, we correct for calcite-aragonite fractionation effects using present-day cave temperatures, as all specimens that needed to be corrected are of late Holocene age. Those speleothems are marked by (**) in table 2. If all speleothems from one cave consist entirely of aragonite (marked (*) in Tab. 2), we performed no correction for fractionation effects. Some isotope samples were referred to as consisting of a 'mixed' mineralogy in SISAL. Those samples were removed from the data set.

#### 3.2.1. Influence of carbonate dissolution conditions

Different pathways of carbonate dissolution affect the carbon isotope composition of drip water (Hendy, 1971; Fohlmeister et al., 2010; Minami et al., 2015). Carbonate dissolution conditions can be similar in the same cave for the same time period, as deduced by radiocarbon reservoir effects of contemporaneously growing speleothems from the same cave (e.g., Fohlmeister et al., 2012; Lechleitner et al., 2016; Cheng et al., 2018; Riechelmann et al., 2019). Furthermore, modelling studies have shown that slight differences in carbonate dissolution conditions have only a small effect on $\delta^{13}C$ values (Hendy, 1971; Fohlmeister et al., 2011; Griffiths et al., 2012). This was recently corroborated by a study on speleothems from Baradla Cave, Hungary (Demény et al., 2017b), where large differences in open to closed dissolution system conditions (10 pmC difference in radiocarbon) had nearly no appreciable effect on $\delta^{13}C$ values.

Here, we further test the influence of carbonate dissolution on $\delta^{13}C$ composition using CaveCalc, a recently developed model for speleothem chemistry and isotopes (Owen et al., 2018). CaveCalc is a numerical model based on PHREEQC, and importantly allows direct quantitative modelling of semi-open dissolution conditions (Owen et al., 2018). Thus, we can test the sensitivity of drip water $\delta^{13}C$ values to system "openness" by varying the volume of soil gas the aqueous solution is in contact with during dissolution (the larger the volume of air, the more open the system).

### 3.2.2. Influence of $CaCO_3$ precipitation processes on $\delta^{13}C$ values

Fractionation effects during degassing of $CO_2$ and carbonate precipitation (including PCP) can lead to substantial variability in speleothem $\delta^{13}C$ values. We assume that temperature, vegetation cover and carbonate dissolution conditions are sufficiently similar for drip sites feeding two or more speleothems from the same cave as long as the cave system is not too large. Thus, we expect the $pCO_2$ in equilibrium with initial drip water to be similar for individual drip locations. Here, initial drip water refers to water that has just achieved $Ca^{2+}$ saturation following dissolution of the carbonate host rock, but where no $CO_2$ has degassed and no carbonate has yet been precipitated. As most stalagmites in our dataset grew in the deep interior of the cave (>100 m distance from any entrance) or close to each other (>5 m distance), when nearer to the entrance, it is also reasonable to assume that the $pCO_2$ level of air is sufficiently constant within the cave. Keeping these assumptions in mind, the only sources of variability in $\delta^{13}C$ data of coevally growing speleothems should be the drip interval and the presence/extent of PCP, the latter defining the degree of supersaturation of $[Ca^{2+}]$ in the solution reaching the top of the speleothem. These two factors also strongly affect growth rate (e.g., Kaufmann, 2003; Mühlinghaus et al., 2007; and Romanov et al., 2008). Once drip water is in equilibrium with cave air $CO_2$ levels, which takes only a few seconds (Dreybrodt and Scholz, 2011; Day and Henderson, 2012), a faster speleothem growth rate requires a shorter drip interval. Furthermore, speleothems with a higher growth rate should have lower $\delta^{13}C$ values as the time for isotopic enrichment of the dissolved inorganic carbon (DIC) and the precipitating $CaCO_3$ is reduced. Thus, we expect a $\delta^{13}C$ offset in contemporaneously growing speleothems with different growth rates. We can quantify this offset by calculating the slope between $\delta^{13}C$ average and average growth rate for two contemporaneously growing speleothems from the same cave. According to the above argument the slope is expected to be negative, given that the initial conditions are similar.

The growth rate - $\delta^{13}C$ offsets obtained from speleothem data were evaluated by first principles of carbonate precipitation and carbon isotope fractionation. We applied a Rayleigh distillation model for $\delta^{13}C$ and an exponential approach for carbonate precipitation, closely following the ISOLUTION modelling approach (Deininger et al., 2012; Deininger and Scholz, 2019). The $Ca^{2+}$ concentration at time t, $[Ca^{2+}](t)$, after the solution is in contact with $CO_2$, decreases progressively and can be approximated by an exponential decay (e.g., Dreybrodt, 1980).

(1) $[Ca^{2+}](t) = ( [Ca^{2+}](t_0) - [Ca^{2+}]_{app}) \exp(-t/\tau) + [Ca^{2+}]_{app}$ in moles/m$^3$

where $[Ca^{2+}](t_0)$ is the initial $Ca^{2+}$ concentration before any carbon is lost from the solution ($t = t_0 = 0$). $[Ca^{2+}]_{app}$ is the $Ca^{2+}$ concentration in equilibrium with cave air $pCO_2$, modulated by inhibiting effects (Dreybrodt et al., 1997; Kaufmann, 2003). The time constant, $\tau$, refers to calcite precipitation rate.

The role of PCP on Eq. 1 is important. While in earlier approaches (Dreybrodt 2008; Mühlinghaus et al., 2009; Scholz et al., 2009; Deininger et al., 2012) carbonate precipitation at the top of a speleothem was considered to be important from $t_0$ to a time point $t_1$, we here use a different approach in order to account for PCP, which is an important process in most caves (e.g., Fairchild et al., 2000; Johnson et al., 2006; Sherwin and Baldini, 2011). In our approach, we define PCP to be acting from $t=t_0$ to $t=t_1$. At $t=t_1$ the droplet is falling onto the stalagmite top and is replaced by the next one at time $t=t_2$ (Fig. 1). Thus, for growth rate calculation purposes, we are only interested in the amount of calcite precipitating between $t_1$ and $t_2$. The amount of precipitated $Ca^{2+}$ (F), between the two time points, $t_1$ and $t_2$, can be calculated by:

(2) $F = ([Ca^{2+}](t_1) - [Ca^{2+}](t_2))$ in moles/m$^3$

The growth rate (GR) of a stalagmite can be determined by F, the typical thickness of the water film layer, R, which is in the order of $10^{-2}$ cm (e.g., Dreybrodt, 1980; Baker et al., 1998), the time interval of precipitation, $t_2$-$t_1$, the density of precipitating calcite, rho (for calcite $\rho$ = 2.689 g/cm3) and the molecular weight, M, of $CaCO_3$ (100.09 g/mol). To express GR in cm/a we have to account for conversion factors (1a is $3.15*10^7$ s and 1 m$^3$ = $10^6$ cm$^3$; Dreybrodt et al., 1980).

(3) $GR = R*F*M/\rho/(t_2-t_1)*3.15*10^7 /10^6$

The temporal evolution of the $\delta^{13}C$ of DIC in drip water ($\delta^{13}C$ (t)) is determined by:

(4) $\delta^{13}C(t) = ( (\delta^{13}C(t_0)/1000 + 1) * ([Ca^{2+}](t)/[Ca^{2+}](t_0))^{\varepsilon} - 1) * 1000$

where $\delta^{13}C$ (t0) is the initial $\delta^{13}C$ of DIC in the drip water and ε is the combined fractionation factor for $^{13}C$, composed from carbon fractionation factors for the transition between $HCO_3^-$ and gaseous $CO_2$ as well as between $HCO_3^-$ and $CaCO_3$ (e.g., Mühlinghaus et al., 2009; Deininger et al., 2012). In order to account for PCP, we do not integrate the $\delta^{13}C$ values of precipitated $CaCO_3$ from $t=t_0$ (the initial drip water) to t=t1 but allow the solution to lose some carbon through degassing and precipitation before the drip impinges on the stalagmite top (Fig. 1). As for the $CaCO_3$ growth rate we focus on the $\delta^{13}C$ value of precipitating $CaCO_3$ at the top of the stalagmite between $t= t_1$ and $t=t_2$ and eq. (4) is transferred to:

(5) $\delta^{13}C(t_2) = ( (\delta^{13}C(t_1)/1000 + 1)* ([Ca^{2+}](t_2)/[Ca^{2+}](t_1))^{\varepsilon} - 1) * 1000$

Therefore, the $\delta^{13}C$ composition of $CaCO_3$ precipitating at the stalagmite top (between $t_1$ and $t_2$, Fig. 1) is the weighted mean of the $\delta^{13}C$ of precipitating $CaCO_3$ and the amount of precipitated $CaCO_3$:

(6) $\delta_{13}C_{CaCO3}(t_1,t_2) = \Sigma\, \delta^{13}C(t'_i)*F(t'_i) / \Sigma\, F(t'_i)$,

where $t'_i = [t_1, t_1+1, t_1+2, ..., t_2]$.

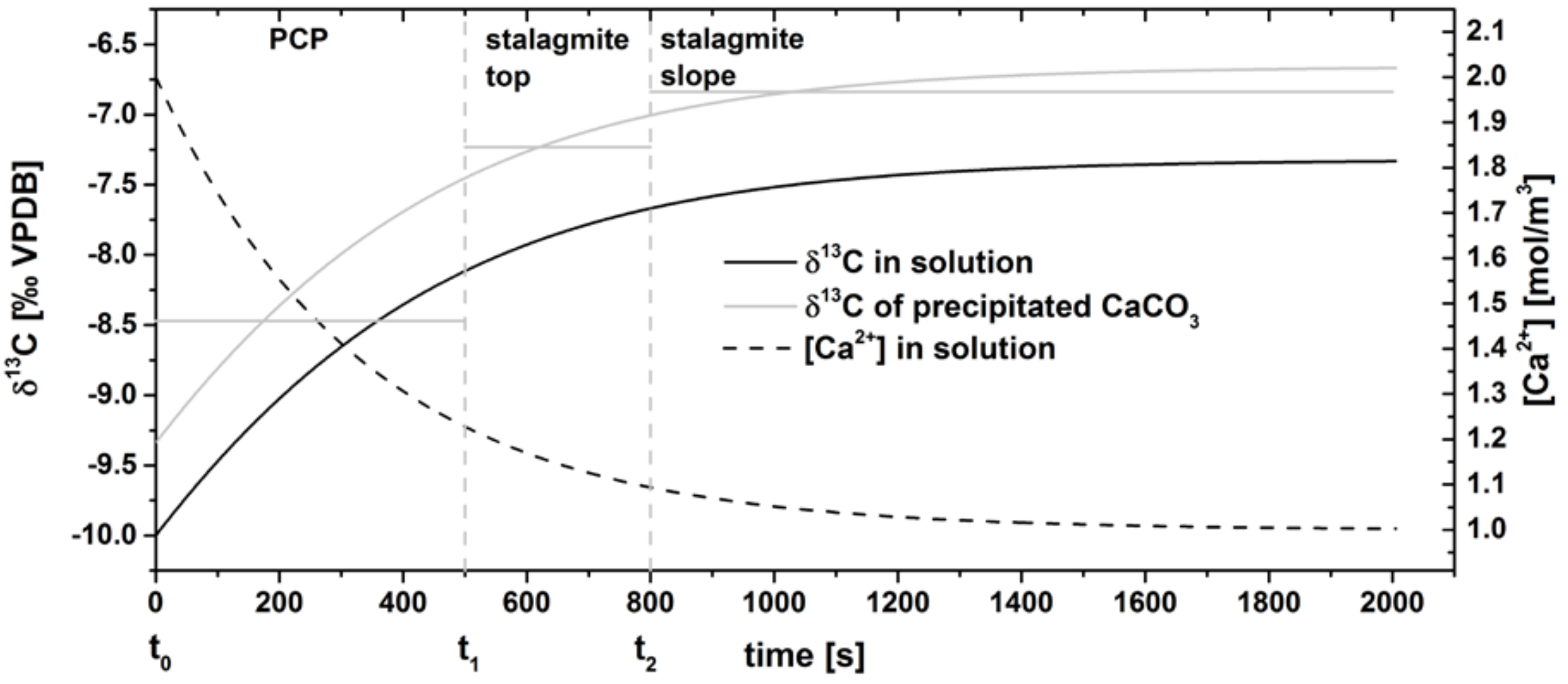

***Fig. 1:*** *Evolution of $\delta^{13}C$ and $[Ca^{2+}]$ in drip water and the respective $\delta^{13}C$ composition of precipitated $CaCO_3$ at 20°C. Between $t_0$ and $t_1$, the water evolves before the droplet reaches the stalagmite top (PCP). Between $t_1$ and $t_2$ (dashed grey lines) carbonate precipitation at the top of the speleothem occurs. After $t_2$, carbonate precipitation takes place along the slopes of the stalagmite, when the drop has been replaced by a new one. The horizontal solid grey lines indicate the average $\delta^{13}C$ isotopic composition of $CaCO_3$ at those three stages.*

## 4. Results

### 4.1. Post-1900 CE speleothem $\delta^{13}C$ data

The speleothems in our dataset are from cave sites with an annual average air temperature ranging from 1 to 27°C, average annual precipitation between 100 and 3000 mm, and altitude between 60 and 3100 m above sea level.

The average $\delta^{13}C$ values of speleothems vary between -12.1 and +3.5 ‰ VPDB. The minimum standard deviation of the average $\delta^{13}C$ is 0.11‰, and the maximum standard deviation is 2.5‰. To check for systematic trends towards lighter or heavier $\delta^{13}C$ values, which would complicate our analysis, we calculated the slope of $\delta^{13}C$ with time for each stalagmite record (Fig. 2A). While the mean of all records is slightly shifted towards a negative slope (-0.0035 ‰/a), the ensemble is nearly Gaussian distributed (Fig. 2B) with a standard deviation of 0.0177. This indicates that applying an ensemble approach can level out various local processes in individual cave environments.

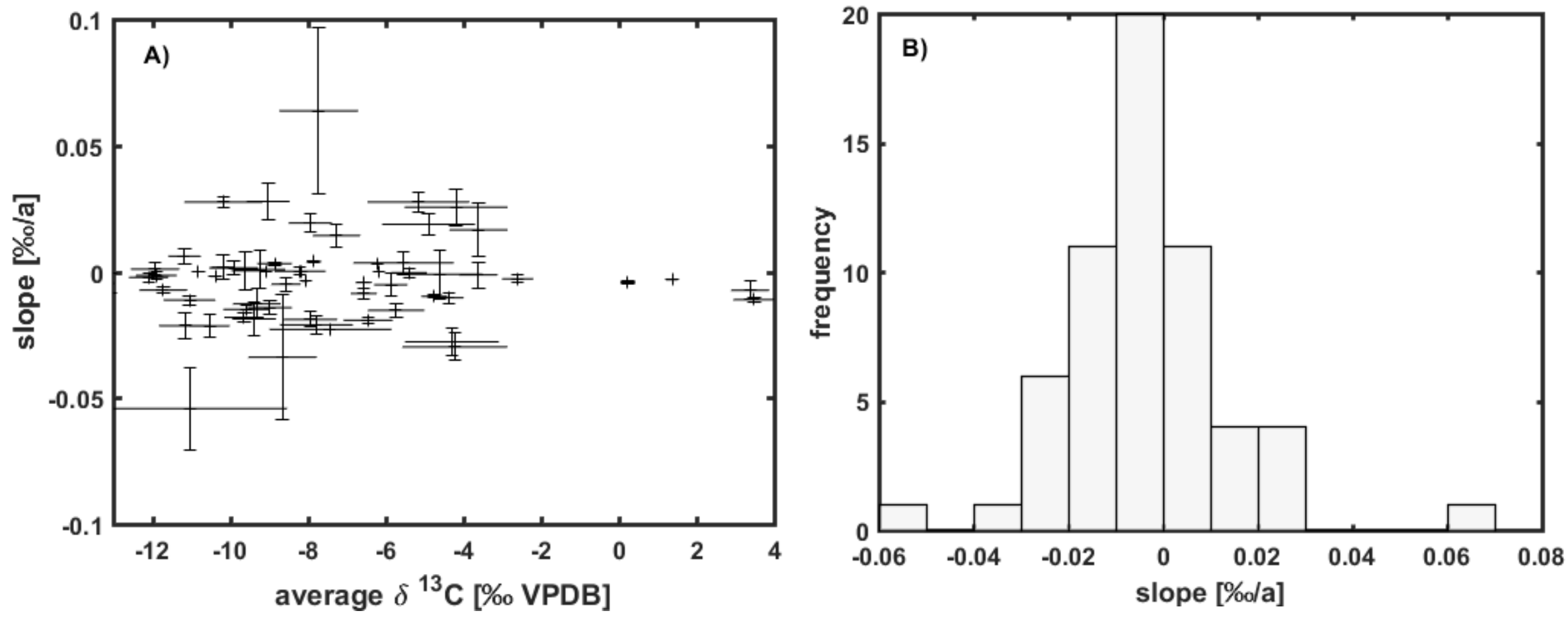

***Fig. 2:*** *A) Slope (*$\delta^{13}C$ */ time) of speleothem* $\delta^{13}C$ values *for the post-1900 CE records in comparison to their average* $\delta^{13}C$ composition*. B) Histogram showing the frequency distribution of the slope for individual speleothems.*

### 4.2. Speleothem $\delta^{13}C$ values from contemporaneously growing speleothems

The average speleothem $\delta^{13}C$ values from this subset vary between ~-13 ‰ and ~+3, and average growth rates vary between ~2*10$^{-5}$ and 1 mm/a (Fig. 3A). The slopes between average $\delta^{13}C$ values and average growth rate of two contemporaneously growing speleothems reveal a bimodal distribution (Fig. 3B). This bimodal distribution is an artefact of the logarithmic scale, which however is necessary to show all data adequately. With the 94 speleothems extracted from the SISAL_v1 database we were able to calculate 76 $\delta^{13}C$ - growth rate slopes between coeval speleothems. 52 out of 76 slopes were negative, i.e., faster growing speleothems have lower $\delta^{13}C$ values, while 24 slopes were positive (Fig. 3A). The positive slopes can vary between 0.2 and 6200 ‰/(mm/a), while negative slopes span a range between -33000 and -0.1 ‰/(mm/a). The frequency maximum of the slopes is between 10 and 100 ‰/(mm/a) for both positive and negative slopes (Fig. 3B).

In the next step, we defined a maximum duration of overlap. If the analysed interval of contemporaneous growth is too long, various environmental and cave parameters may have changed (even if in a similar manner for both speleothems) potentially complicating the interpretation. Thus, we split periods of contemporaneous growth of speleothems into periods of 1000 years. For example, when the duration of contemporaneous growth between two stalagmites is 10000 years, we divided this interval into ten phases and calculated the $\delta^{13}C$-growth rate lines for each of the 10 phases instead of one slope for the entire 10000 year interval. This modification greatly increased the number of calculated slopes ($\delta^{13}C$ vs growth rate, Fig. 3C). Encouragingly, the general $\delta^{13}C$-growth rate relationship (i.e., the slope) is well reproduced even when using this larger dataset (Figs. 3C and 3D), and suggests that these relationships are robust at different timescales.

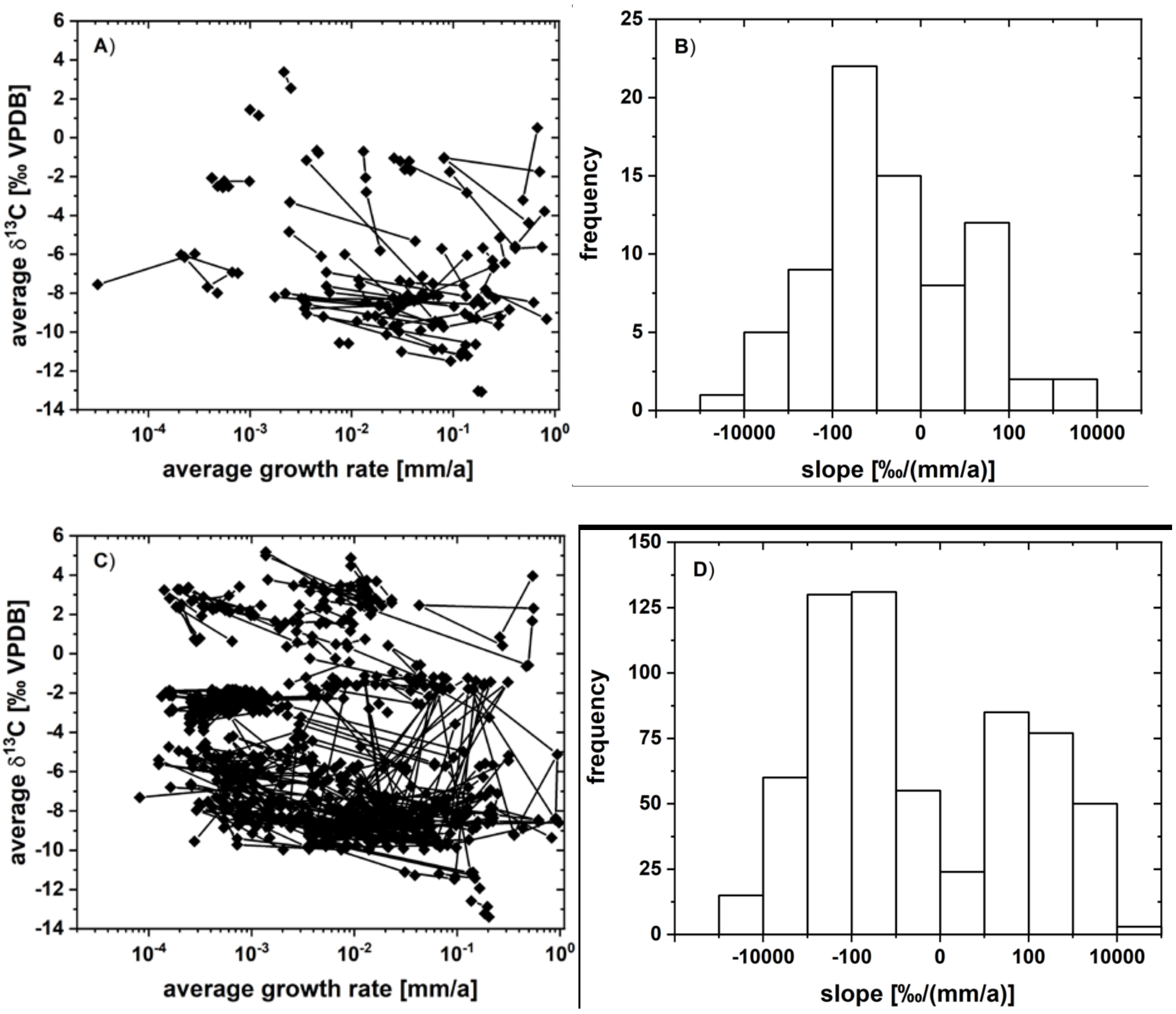

***Fig. 3:*** *A) Average* $\delta^{13}C$ *values and average growth rates for periods of contemporaneous speleothem growth. Speleothem pairs from the same cave are connected by a straight line. Error estimates for growth rate are about 10 % for those records with large age errors. B) Histogram showing the frequency distribution for the slopes (*$\delta^{13}C$ *– growth rate). Please note the logarithmic scale on the x-axes. A stability analysis with using at maximum 1000 year periods (see text for details) suggests that this pattern is robust (C as in A, D as in B only with analysing 1000 year long windows).*

## 5. Discussion

### 5.1. Controls on modern speleothem $\delta^{13}C$ values

As speleothem $\delta^{13}C$ values can have a large biogenic component originating from vegetation and/or soil activity above the cave, we compare the average $\delta^{13}C$ values of modern speleothems (post-1900 CE) with instrumental climate and vegetation data. Although it is possible to convert each category of vegetation from the colour code on the map into numbers, defining the correlation factors or other statistical quantification of the relationship between vegetation cover and speleothem $\delta^{13}C$ values is less reliable. The map resolution is too coarse to provide accurate correlation with the local vegetation at the cave site, and although some sites have better description of the local type of vegetation cover, the interpretation may be biased due to the limited information from the other sites where complete vegetation description of the cave surrounding is lacking. Hence, we use general observations by comparing the vegetation zones and speleothem $\delta^{13}C$ values, which were also colour coded for every per mil change (Fig. 4). This qualitative observation suggests that higher speleothem $\delta^{13}C$ values tend to be associated with vegetation zones that are generally characterized by less vegetation, and vice versa, and it could be used as a general framework to set the baseline and range of $\delta^{13}C$ values for modern speleothems. Nevertheless, caution should be taken when interpreting $\delta^{13}C$ in regions where vegetation cover is more heterogeneous.

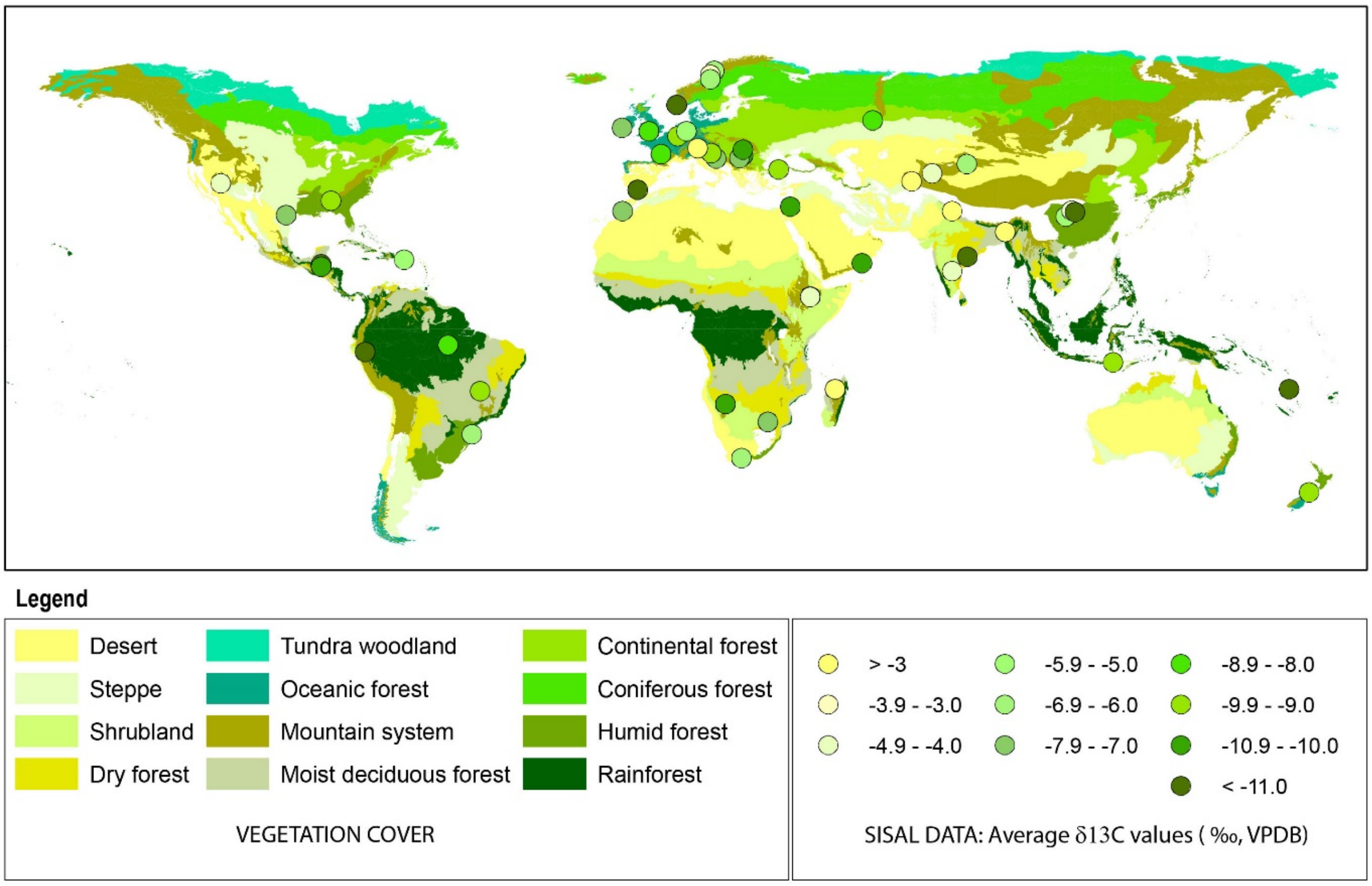

***Fig. 4:*** *Global map showing the relationship between vegetation cover and average speleothem* $\delta^{13}$C *data post-1900 CE. Vegetation map is from Global Land Cover 2000 Project (GLC2000) based on SPOT-VEGETATION satellite imagery (Aaron and Gibbs, 2008).*

Because vegetation type and density strongly depend on the amount of precipitation, site altitude and local air temperature, we also assess the relationships between speleothem $\delta^{13}$C averages and these three variables. First we focus on altitude and annual precipitation (Fig. 5).

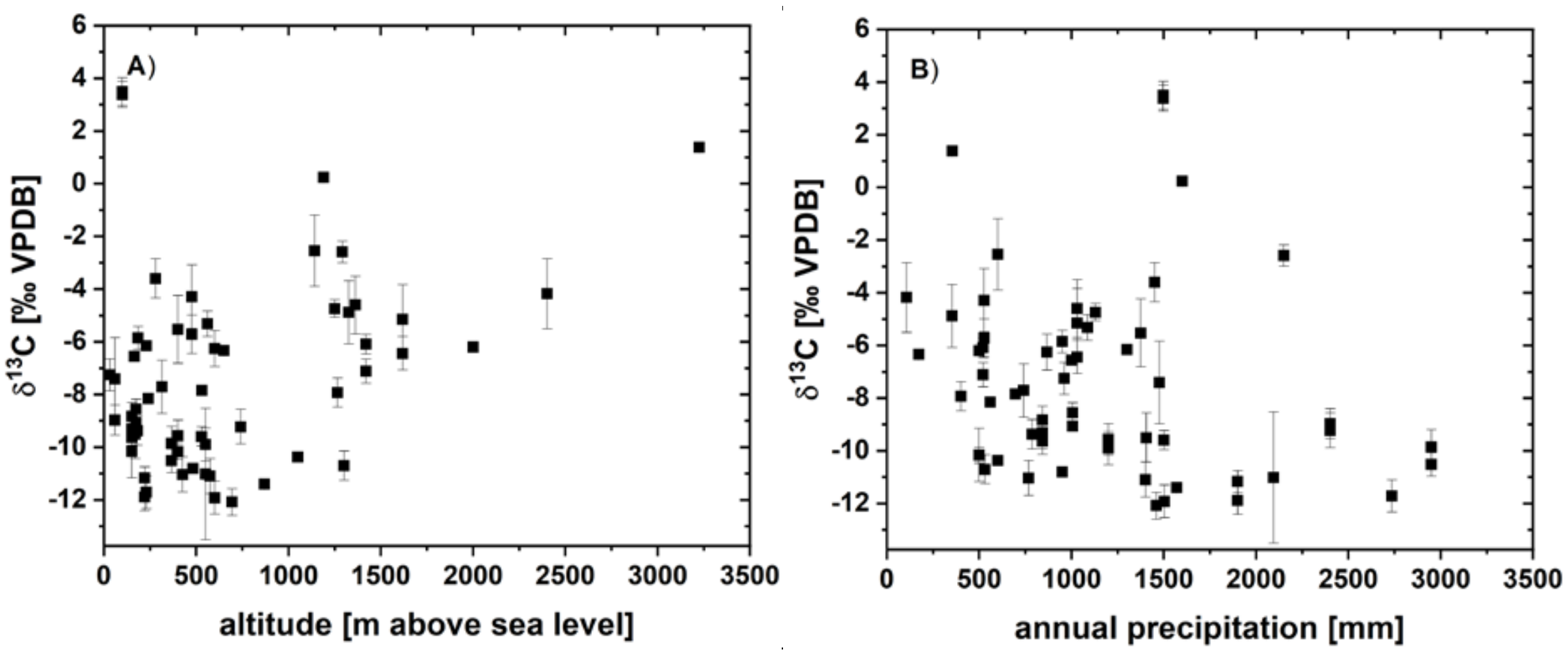

***Fig. 5:*** *Average speleothem* $\delta^{13}$C data *post-1900 CE vs altitude of the cave (A) and vs amount of precipitation (B).*

With respect to altitude, there appears to be no significant relationship with speleothem $\delta^{13}C$ data. When using the Spearman rank correlation, which better assesses monotonic relationships in general and is thus less sensitive to outliers, a correlation coefficient of 0.2 with p = 0.13 is obtained. Nevertheless, the marked absence of low $\delta^{13}C$ values at high-altitude sites is interesting (Fig. 5A). These results agree with findings from a locally confined set of caves, monitored above an altitude gradient in the southern European Alps (Johnston et al., 2013). In that study, lower speleothem $\delta^{13}C$ values are not found in high-altitude caves, while at lower altitudes, the entire range of $\delta^{13}C$ values is present. This speleothem $\delta^{13}C$ behaviour with respect to altitude may be related to vegetation cover and/or soil thickness, which becomes sparser at higher altitudes. Other cave site specific aspects like drip interval or cave ventilation are viewed as having minimal influence within such a large set of data. A relationship between $\delta^{13}C$ and altitude is also suggested by a principal component analysis (PCA, Fig. S1), where the $\delta^{13}C$ values and altitude vectors point approximately to the same direction.

We found a weak but significant negative relationship between speleothem $\delta^{13}C$ values and the amount of precipitation ($\rho$ = -0.27, p = 0.04, Fig. 5B). This is also consistent with the PCA results where $\delta^{13}C$ and precipitation show contrasting behaviour on PC1 (Fig. S1). The full range of speleothem $\delta^{13}C$ values is observed for caves located in regions with annual precipitation below 1500 mm. Another remarkable feature is that high $\delta^{13}C$ values (>-8 ‰) are not found in regions receiving more than ~1500 mm of precipitation per year. From this observation, we deduce that high amounts of precipitation favour the development of high-density vegetation, which promotes lower speleothem $\delta^{13}C$ values. The only exception is Wah Shikhar Cave with a mean $\delta^{13}C$ value of ~-2.6 and an average annual precipitation of ~2150 mm/a (Sinha et al., 2011). Although the area above Wah Shikhar Cave is also densely vegetated (Sinha et al., 2011), processes in the cave and karst must have strong influence on speleothem $\delta^{13}C$ values, favouring the transition from the light isotopic soil $CO_2$ values towards heavy isotopic values in the speleothem.

## 5.2. Temperature and speleothem $\delta^{13}C$

The most interesting and complex pattern evolves when we compare the $\delta^{13}C$ averages of speleothems in the post 1900 CE era with the present day mean annual temperature. Although there is no clear relationship in the data ($\rho$ = -0.07, p = 0.62), as also confirmed by the PCA (Fig. S1), temperature has the potential to be responsible for the overall sensitivity of speleothem $\delta^{13}C$ to vegetation and climate conditions. As photosynthesis and respiration depend on temperature, we expect differences in the $\delta^{13}C$ values of speleothems from cold and warm/hot sites (indicated as tracks A and B in Fig. 6). From this general pattern several cave specific processes can contribute to a deviation from this relationship and might be summarised by three categories in our dataset (labelled #1, #2, #3 in Fig. 6). Four out of these five general clusters are also well identified (Fig. S3) by an unsupervised machine learning algorithm (DBSCAN; Ester et al., 1996), suggesting that our grouping is objective. Details and results of this algorithm are provided in the supplement.

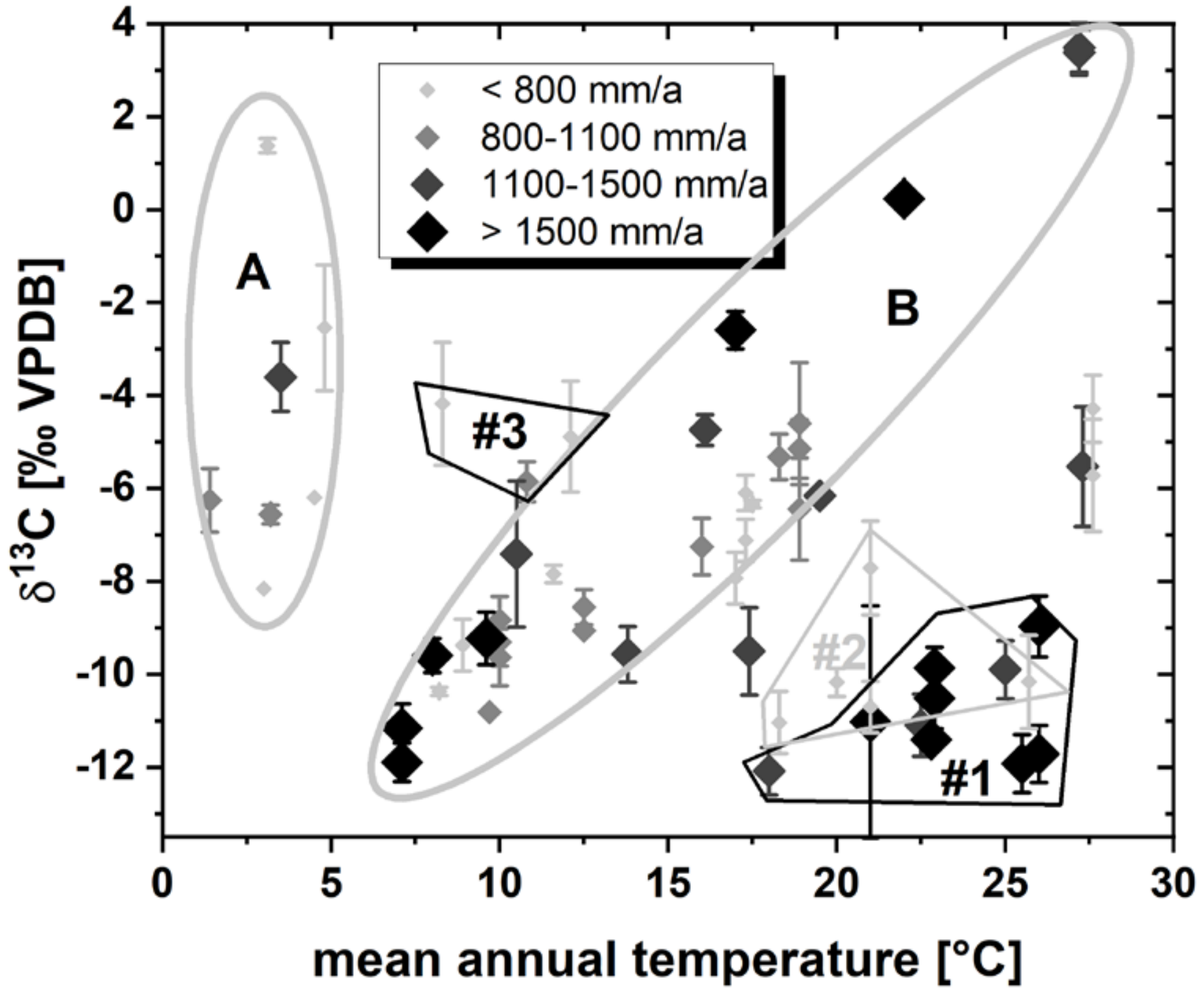

***Fig. 6:*** *Clustering pattern between* $\delta^{13}C$ averages *and mean annual temperature (diamonds), highlighting the highly non-linear relationship between both parameters. Grey-tone and size of the diamonds refer to the amount of annual precipitation at the cave location. The two near-linear branches (A and B), discussed in the text, for low and high temperatures are indicated by the grey lines (labelled A and B). The three marked clusters (#1, #2 and #3) highlight special characteristics of those speleothem* $\delta^{13}C$ *values. See section 5.2 for details.*

For the cold temperature branch (labelled A in Fig. 6), one possible explanation could be the production of soil gas $CO_2$. The soil gas $\delta^{13}C$ composition is very sensitive to diffusion of $CO_2$ into soils when soil respiration rates are small and the atmospheric component becomes more important (Cerling et al., 1984). Soil respiration rates between ~ 0 and 1 mmoles/m^2/hr can easily explain $\delta^{13}C$ variations of about 12 ‰ in soil gas $CO_2$ (between -21 and -9‰ for soil respiration rates between 0 and 1 mmoles/m^2/hr). At low temperatures, soil respiration rates are generally lower than under warmer temperatures for similar vegetation cover (Raich and Schlesinger, 1992; Raich and Potter, 1995; Klätterer et al., 1999). For comparison, soil respiration rates for grassland soils during the growing season can reach values between 6 and 9 mmoles/m^2/hr, while during the dry or cool non-growing season they are typically about 1 mmoles/m^2/hr (Singh and Gupta, 1977; Schlesinger, 1977; Parker et al., 1983). At freezing conditions, soil respiration rates can drop close to 0 mmoles/m^2/hr (Kucera and Kirkham 1971).

Thus, it appears likely that diffusion of atmospheric $CO_2$ into soils at low soil respiration rate sites can explain the large spread of $\delta^{13}C$ data we observe for low temperature sites (below 5°C MAT; Fig. 6). Speleothems from these caves show a wide $\delta^{13}C$ range, varying from -8 to +2 ‰, despite the small sample size (i.e., only seven speleothems were available for evaluation). This group has no significant correlation to temperature, but roughly follow the expected trend. The average speleothem $\delta^{13}C$ values at these cold sites are higher than those at slightly warmer caves (around 7-10 °C).

Speleothems following the warm temperature branch (~7 to 27 °C, Fig. 6), also show a wide range in $\delta^{13}C$ values (between -~12 and +3 ‰). With increasing mean annual temperatures, $\delta^{13}C$ values also increase, following an approximately linear relationship. Optimal conditions for speleothem growth appear to be between ~7 and 15 °C, where most speleothems cluster, with low $\delta^{13}C$ values (-12 to -8 ‰) that suggest a large imprint from soil respired $CO_2$. One possible mechanism to explain the high $\delta^{13}C$ values (-2 to +4 ‰) at high temperatures is increasing heat and drought stress on vegetation. As opposed to soils from cold regions, those from higher temperature regions (>15 °C) can be affected by enhanced evaporation, which can quickly reduce the water stored in soils, which in turn is responsible for changing fractionation strength during photosynthesis (Bowling et al., 2002; Hartman and Danin, 2010; Buchmann et al., 1996). Temperature can thus be responsible for reduced vegetation density and soil respiration rates, even when the overall amount of precipitation is high. This in turn affects the soil gas $\delta^{13}C$ via the earlier discussed fractionation effects during $CO_2$ diffusion out of the soil into the free atmosphere. Another reason for increased speleothem $\delta^{13}C$ values might be that C4 type plants become more likely at higher temperatures as seen at Anjohibe Cave (Madagascar), which is nearly completely covered by C4 plants (average $\delta^{13}C$ of stalagmites AB2 and AB3 = +3.4 and +3.5 ‰; MAT = 27.2 °C; Burns et al., 2016; Scroxton et al., 2017; Voarintsoa et al., 2017c). As the C4 metabolic pathway fractionates the carbon isotopes less strongly, a higher proportion of C4 plants can also explain the trend towards higher speleothem $\delta^{13}C$ values at high temperatures.

We observe significant deviations in speleothem $\delta^{13}C$ values from the warm temperature branch. One category of data (region '#1') forms a cluster with very negative $\delta^{13}C$ values (around –12‰) at high temperature sites (Fig. 6). These samples are from climates experiencing strong precipitation seasonality, e.g., monsoonal, and high amounts of annual precipitation as in Jhumar Cave, India (Sinha et al., 2011), Liang Luar Cave, Indonesia (Griffiths et al., 2016) or Taurius Cave, Vanuatu (Partin et al., 2013). In this case, it is likely that despite the high temperatures reigning at the sites, enough water is available to maintain a dense vegetation, which translates to high soil respiration rates, and low $\delta^{13}C$ values in soil gas and speleothems.

Of special interest are the five speleothem $\delta^{13}C$ data points in region '#2' (Fig. 6), which largely overlap with region #1. These samples have low $\delta^{13}C$ values (-7.5 to -11 ‰) and high temperatures, but are affected by a much more arid climate than the samples in region '#1'. Two of the five samples are from St. Michaels Cave, Gibraltar (Mattey et al., 2008) and Natural Bridge Caverns, Texas (Wong et al., 2015). These two sites are well monitored and their dominant cave processes and carbon fluxes are well understood (Mattey et al., 2016; Breecker et al., 2012, Meyer et al., 2014; Bergel et al., 2017). For both caves, it has been recognised that a deep, $^{13}C$ depleted carbon source in the karst must exist in addition to the surface soil $CO_2$. This would explain the relatively low speleothem $\delta^{13}C$ values at these sites, despite the warm and arid local conditions and relatively sparse vegetation at present. The other three data points are from Dante Cave, Namibia (Voarintsoa et al. 2017a), Soreq Cave, Israel (Bar-Matthews et al., 2003) and Defore Cave, Oman (Burns et al., 2002). Unfortunately, no detailed monitoring data of carbon transfer dynamics is hitherto available for these three caves, and therefore we cannot test whether they are also affected by the presence of deep carbon reservoirs. This proves how important cave monitoring is for understanding individual proxy time series but also to understand data in a more general context like this compilation.

Finally, the cluster in region '#3' (Fig. 6) is composed of three speleothems characterised by unusually high $\delta^{13}C$ values compared to other speleothems in the same temperature range. These three speleothems are from Bunker Cave, Germany (Fohlmeister et al., 2012), Bir-Uja Cave, Kyrgyzstan (Fohlmeister et al., 2017), and Leviathan Cave, Nevada (Lachniet et al., 2014). Bunker Cave was artificially opened after its discovery in 1860 CE, which likely lead to stronger ventilation and an increase in $\delta^{13}C$ values (Riechelmann et al.,2011; Fohlmeister et al., 2012). Speleothem values prior to 1860 CE average around -9 ‰ (instead of -5.9‰ in the period after 1860 CE), which better matches predictions based on vegetation cover (Fig. 4) and temperature (Fig. 6). A similar explanation is provided for the Bir-Uja Cave stalagmite, where fractionation processes within a small cave with a large opening are dominant (Fohlmeister et al, 2017). Such strong fractionation effects are often reported for well ventilated caves (Spötl et al., 2005; Frisia et al., 2011; Tremaine et al., 2011). Little information about fractionation processes is available for Leviathan Cave, therefore we cannot make final statements about the reason for its relatively high $\delta^{13}C$ values. However, site descriptions suggest that vegetation cover at this site is sparse and dominated by grasses, as expected for the very arid conditions in the Great Basin (Lachniet et al., 2014). This suggests that again, the low soil respiration rate at this high elevation site (2400 m above sea level) might be contributing to the high average $\delta^{13}C$ values.

### 5.3. Governing processes in karst and speleothem $\delta^{13}C$ values

In the previous subsections we analysed the influence of vegetation, temperature, precipitation and altitude on carbonate $\delta^{13}C$ values. In order to eliminate these site-specific external factors from our analysis, we compare $\delta^{13}C$ values of contemporaneously grown speleothems from the same cave. Thus, any differences in the average $\delta^{13}C$ values of contemporaneously grown speleothems can be attributed to changes in the type of host rock dissolution and fractionation processes during PCP or $CO_2$ degassing.

In this section and in section 5.4 we will focus on the direction and steepness of the slope between the average $\delta^{13}C$ value and the average growth rate of two contemporaneously growing speleothems (Fig. 3A). The detected bimodal distribution in the frequency of the slopes (Fig. 3B) is surprising under the given boundary conditions of similar drip water and cave air characteristics. More than 2/3 of all analysed speleothem pairs show a negative slope between $\delta^{13}C$ and growth rate (as expected from previous modelling studies, e.g., Mühlinghaus et al., 2007; Romanov et al., 2008; Dreybrodt and Scholz 2011 and the description in Sec. 3.2.2), while the remaining show a positive slope. This is an interesting result and raises the question for the driving mechanism behind these relationships.

First, we evaluate the possibility that carbonate dissolution processes can explain the observed behaviour in the slopes of $\delta^{13}C$ values vs. growth rates. For this purpose, we use the forward model CaveCalc (Owen et al., 2018) to model the evolution of $\delta^{13}C$ during carbonate dissolution processes in the karst. In this model the degree of open to closed carbonate dissolution conditions is represented by an adjustable amount of soil gas, which is able to supply $CO_2$ during the dissolution of carbonate bed rock and contribute $CO_2$ for the exchange of carbon between the dissolved carbon species and the specified gas volume. The larger the specified gas volume, the more open the carbonate dissolution conditions.

For the modelling, we assumed soil gas $\delta^{13}C$ values of -25 ‰ with a concentration of 10,000 ppm, which are typical values for soils (e.g., Spötl et al., 2005, Frisia et al., 2011, Mattey et al., 2016). We specify a calcitic bedrock and a temperature of 10°C. While the previous parameters were kept constant, we varied the amount of gas volume (between 1 and 500 L) that is in contact with the acidic solution during the dissolution of the host rock (Fig. 7), to simulate changes in the carbonate dissolution system (open vs closed conditions).

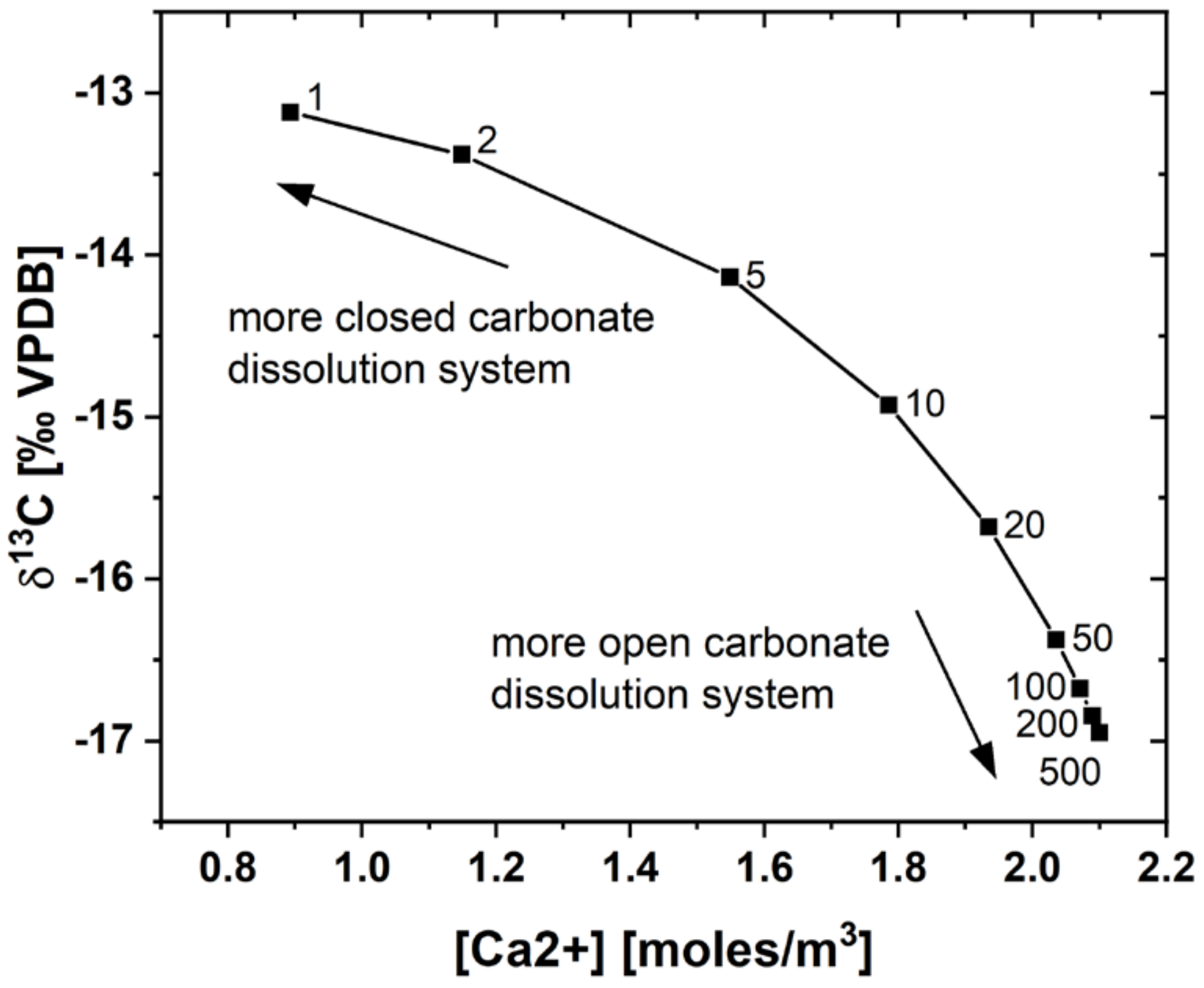

***Fig. 7:*** *Carbonate dissolution under various degrees of open/closed conditions using CaveCalc (Owen et al., 2018). Numbers refer to the volume of gas (in l) in contact with the water during dissolution. The more open the carbonate dissolution system (larger gas volume), the higher [*$Ca^{2+}$*] and the lower the* $\delta^{13}C$ *of DIC of the [*$Ca^{2+}$*] saturated drip water. The resulting range of slopes (*$\delta^{13}C$ *vs growth rate) is between -6 and -61 ‰/(mm/a).*

Under more open carbonate dissolution systems, e.g. when the gas volume is larger, $\delta^{13}C$ values decrease and [$Ca^{2+}$] increases. This is because more soil $CO_2$ is available for bedrock dissolution, leading to more $CaCO_3$ dissolution (increase in [$Ca^{2+}$]), while providing a large reservoir of low - $\delta^{13}C$ air with which the solution can re-equilibrate. The relationship between $\delta^{13}C$ values and [$Ca^{2+}$] is non-linear with a stronger increase in [$Ca^{2+}$] than a decrease in $\delta^{13}C$ values for small gas volumes (near completely closed system). For large gas volumes, [$Ca^{2+}$] increases more slowly than for small gas volumes, while $\delta^{13}C$ decreases more rapidly. This behaviour can be explained by the highly nonlinear dissolution of $CaCO_3$ with respect to the available gaseous $CO_2$ (Dreybrodt, 1988). The [$Ca^{2+}$] - $\delta^{13}C$ relationship found by CaveCalc is in agreement with an alternative carbon isotope enabled karst dissolution model (Fohlmeister et al., 2011).

Higher [$Ca^{2+}$] in the solution entering the cave favours higher growth rates. If we assume for simplicity that both drip sites have similar drip intervals, which is also an important factor for growth rate and $\delta^{13}C$ evolution (see Sec. 5.4), a negative slope between $\delta^{13}C$ values and growth rate is established by a variation in dissolution conditions, i.e., a more open system dissolution leads to lower $\delta^{13}C$ values and higher [$Ca^{2+}$] and thus higher growth rates, and vice versa for more closed system conditions. This is in line with the majority of the $\delta^{13}C$ - growth rate slopes found in contemporaneously growing speleothems from the same cave (ca. two thirds of datasets have negative slope, Fig. 3B). To test our assumption, we use the relationship between $\delta^{13}C$ values and [$Ca^{2+}$] to calculate the slope expected from the simulated dissolution regimes. The concentration of $Ca^{2+}$ can be transferred to growth rates by applying equation (3). If we assume a typical cave air $pCO_2$ of 500ppm, the equilibrium [$Ca^{2+}$] is about $0.77*10^{-3}$ moles/l. Furthermore, we assume there is one drip per second for both sites. We choose this short interval in order to only focus on the effect of $CaCO_3$ dissolution. The effect of carbonate precipitation under different drip intervals is discussed later (Sec. 5.4 and 5.5). This allows us to calculate the precipitation rate. The extreme values for the slope are -6 ‰/(mm/a) for near closed dissolution conditions (gas volume = 1 and 2 l) and -63 ‰/(mm/a) for close to completely open dissolution conditions (gas volume = 200 and 500 l). Those values fall very closely to the frequency maximum of the observed frequency distribution for the branch with the negative slope (Fig. 3B). Thus, the steepest negative slopes (-200 - -1000‰/(mm/a), Fig 3B) calculated from the SISAL_v1 dataset cannot be explained by carbonate dissolution systematics. Furthermore, variations in dissolution systematics can also not explain the positive slopes obtained in about one third of cases in our dataset. Thus, the positive slope must be related to processes occurring during degassing of $CO_2$ and $CaCO_3$ precipitation in the cave after the dissolution of $CaCO_3$ is completed.

## 5.4. Governing processes on speleothem $\delta^{13}C$ values during $CO_2$ degassing and $CaCO_3$ precipitation

Studies focusing on radiocarbon analysis of contemporaneously growing speleothems have shown that variations in the open/closed ratio are often small between speleothems from the same cave (Lechleitner et al., 2016, Demeny et al., 2017, Riechelmann et al., 2019; Markowska et al., 2019). This requires that initial $\delta^{13}C$ values and $[Ca^{2+}]$ are approximately equal for two drip sites when the water just reaches $Ca^{2+}$-saturation. Where there are small differences in the open-closed carbonate dissolution system for two drip sites, we have shown that $\delta^{13}C$ and $[Ca^{2+}]$ of the Ca-saturated solution is not expected to change significantly (Fig 7).

When considering the effects of $CO_2$ degassing and $CaCO_3$ precipitation, the only variable parameters are the time interval during which the water is in contact with cave air before reaching the top of the speleothem and the time interval during which the drop is on the top of the speleothem before it is replaced by the next incoming drop. The first time period essentially falls within the time the drip water solution is affected by PCP. Its duration will influence the isotopic composition and growth rate of the speleothem. For our analysis, we first assume that no PCP occurs, e.g., water is not in contact with air prior to reaching the top of the stalagmite (Fig. 8). To model the resulting slopes in two contemporaneously growing speleothems with different drip intervals, we simulate the chemical evolution of two drip sites, by systematically changing cave temperature (T = 5, 10 and 20°C), $[Ca^{2+}]$ in drip water (equivalent to a cave air $pCO_2$ = 500 and 1000 ppm) in the same way for both speleothems. Additionally, we change the drip interval difference between the two speleothems, and calculate how these factors will affect the slope between two contemporaneously growing speleothems using a Rayleigh model (Sec. 3.2.2).

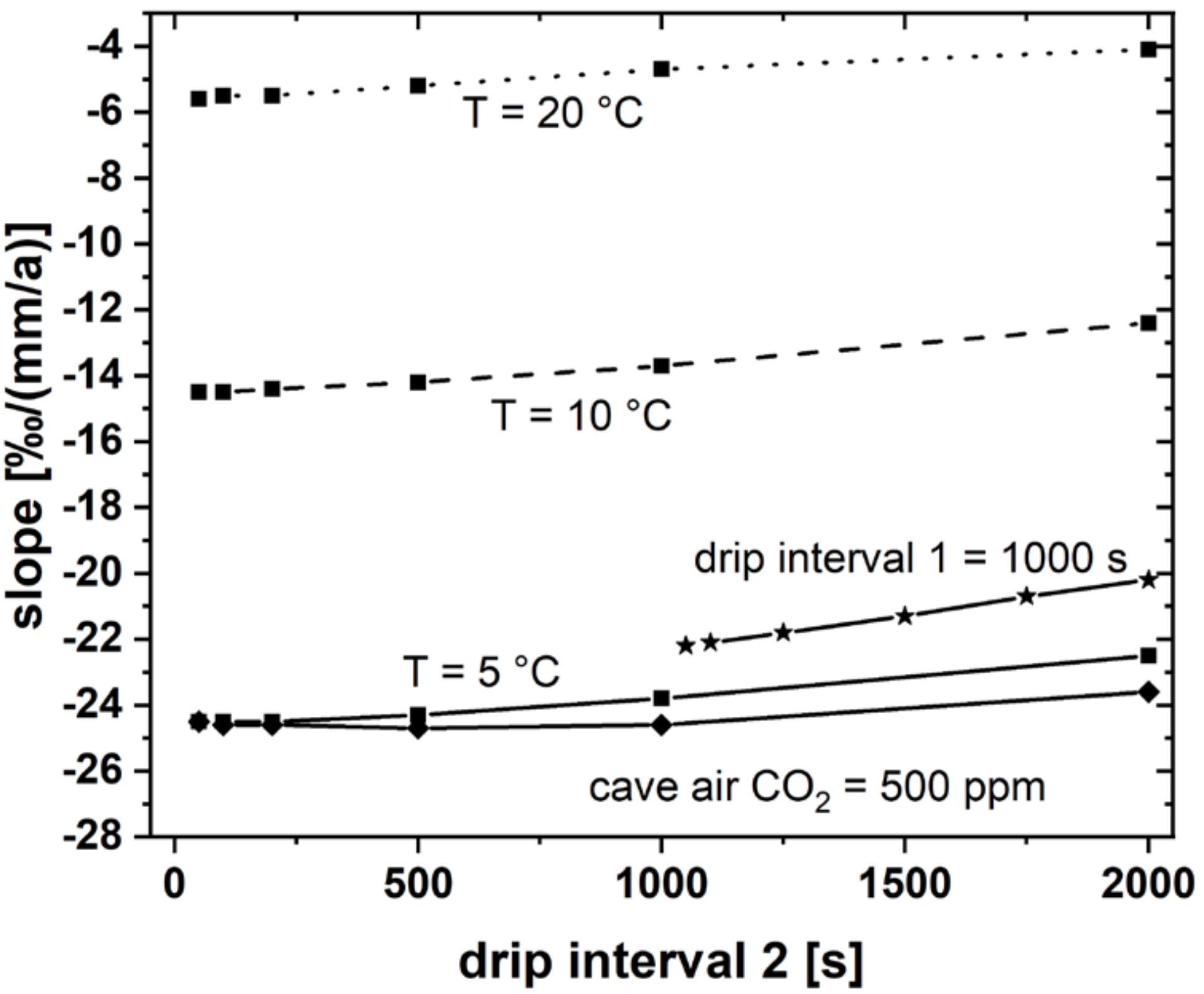

***Fig. 8:*** *The slope of* $\delta^{13}C$ *over growth rate for two contemporaneously growing speleothems over drip interval variations of the second speleothem. When not explicitly modelled, T is set to 5 °C, cave air* $pCO_2$ *to 1000 ppm and drip interval of the first speleothem to 1 s. Initial drip water* $\delta^{13}C$ composition *is -10 ‰ and* [$Ca^{2+}$] *is 2 moles/m^3, which corresponds to a solution in equilibrium with approximately 5000 ppm of* $CO_2$.

The Rayleigh model predicts that the greater the drip interval difference between the two speleothems, the less negative the slope (Fig. 8). If both drip intervals are long (i.e., both are above 1000 s), the slope also becomes less negative. Both observations can be explained by the different rate of change for $\delta^{13}C$ values and [$Ca^{2+}$] during $CaCO_3$ precipitation. The used parameters for calcite precipitation rate, $\tau$ (Eq. 1) and for fractionation factors, $\varepsilon$ (Eq. 4) require that $\delta^{13}C$ values of DIC reaches equilibrium faster than [$Ca^{2+}$] (Fig. 1). This behaviour is even more pronounced in the $\delta^{13}C$ evolution of precipitated $CaCO_3$ and growth rate (supplementary Fig. S4). Thus, under the scenario of a large difference in drip intervals between two speleothems, the difference between DIC $\delta^{13}C$ values at different time steps increases more slowly than that of [$Ca^{2+}$] (and therefore growth rate), resulting in a smaller absolute value of the slope. When the oversaturation of the [$Ca^{2+}$] saturated solution is larger (i.e., cave air $CO_2$ is 500 instead of 1000 ppm) the slope will only be slightly more negative. The largest effect on slope magnitude results from temperature variations. Increasing temperatures lead to less negative slopes, and the effect is stronger at lower T. An explanation for this behaviour is provided by the temperature-dependent parameters $\tau$ and $\varepsilon$, which lead to a more pronounced difference in the rate of change for $\delta^{13}C$ values in DIC and [$Ca^{2+}$] during degassing of $CO_2$ and precipitation of $CaCO_3$. As for the carbonate dissolution conditions (Sec. 5.3), the slope is always negative and fits well to the majority of the slopes analysed for the speleothems extracted from SISAL (between -1 to -100 ‰/(mm/a) – see Fig. 3B).

However, this process fails to provide a mechanism for the observed positive slopes in about one third of the data set on contemporaneously growing speleothems, suggesting that another factor must be at play to generate this behaviour. Thus, we allow for a variable period of PCP in our model experiments but keep the duration of PCP at equal length for both speleothems. In addition, temperature and cave air $pCO_2$ were kept constant (Fig. 9).

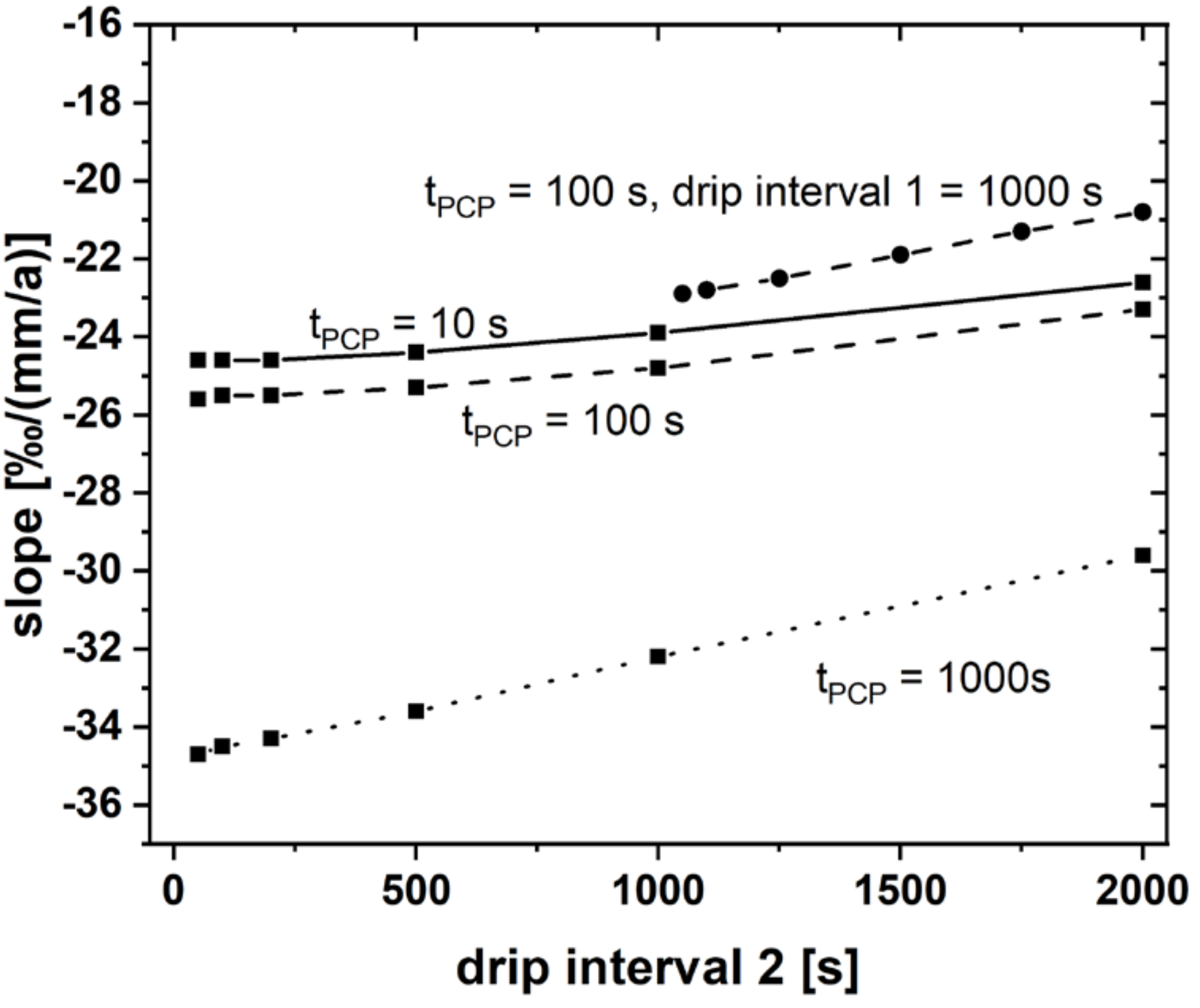

***Fig. 9:*** *The slope of* $\delta^{13}C$ *over growth rate for two contemporaneously growing speleothems over drip interval variations of the second speleothem. In contrast to Fig. 8, PCP is enabled and the time for PCP is varied (but both speleothems experience equal PCP duration). Temperature is set to 5 °C, cave air* $pCO_2$ *to 1000 ppm and when not explicitly modelled drip interval of the first speleothem to 1 s. Initial drip water* $\delta^{13}C$ composition *is -10 ‰ and* [$Ca^{2+}$] *is 2 moles/m^3, which corresponds to a solution in equilibrium with approximately 5 000 ppm of* $CO_2$*.*

The main features of the slope with respect to drip intervals (Fig. 8) remain valid even when accounting for PCP. We observe that the slope in $\delta^{13}C$ values vs growth rate becomes more negative with increasing time for PCP. It is likely that the different rate of change in the evolution of cave drip water with respect to $\delta^{13}C$ and [$Ca^{2+}$] is again responsible for this behaviour. While this analysis also reveals slopes that are in agreement with most of our analysed $\delta^{13}C$ data, it remains impossible to produce the observed positive slopes.

Therefore, in the next step we applied a more comprehensive approach, which likely reflects more general conditions in a cave system, by allowing individual variability in the duration for PCP at the two drip sites. First, we investigate the slopes when the time for PCP is longer for drip site 2 compared to drip site 1 and when drip site 2 also has a longer drip interval than drip site 1. This approach produces similar relationships as compared to results from an equal time for PCP for both drip locations, and thus cannot explain positive slopes (Figs. 8, 9). We then prescribed longer duration for PCP for drip site 1 than for drip site 2, but still leaving the drip interval of site 1 shorter than for site 2 (Fig. 10). These conditions could reflect a drip site 1 where water is in contact with cave air $pCO_2$ for a prolonged period of time, e.g. water running down a cascade of cave-roof carbonates, and a drip site 2 where a slowly dripping soda straw would result in only a short exposure of drip water to cave air $CO_2$.

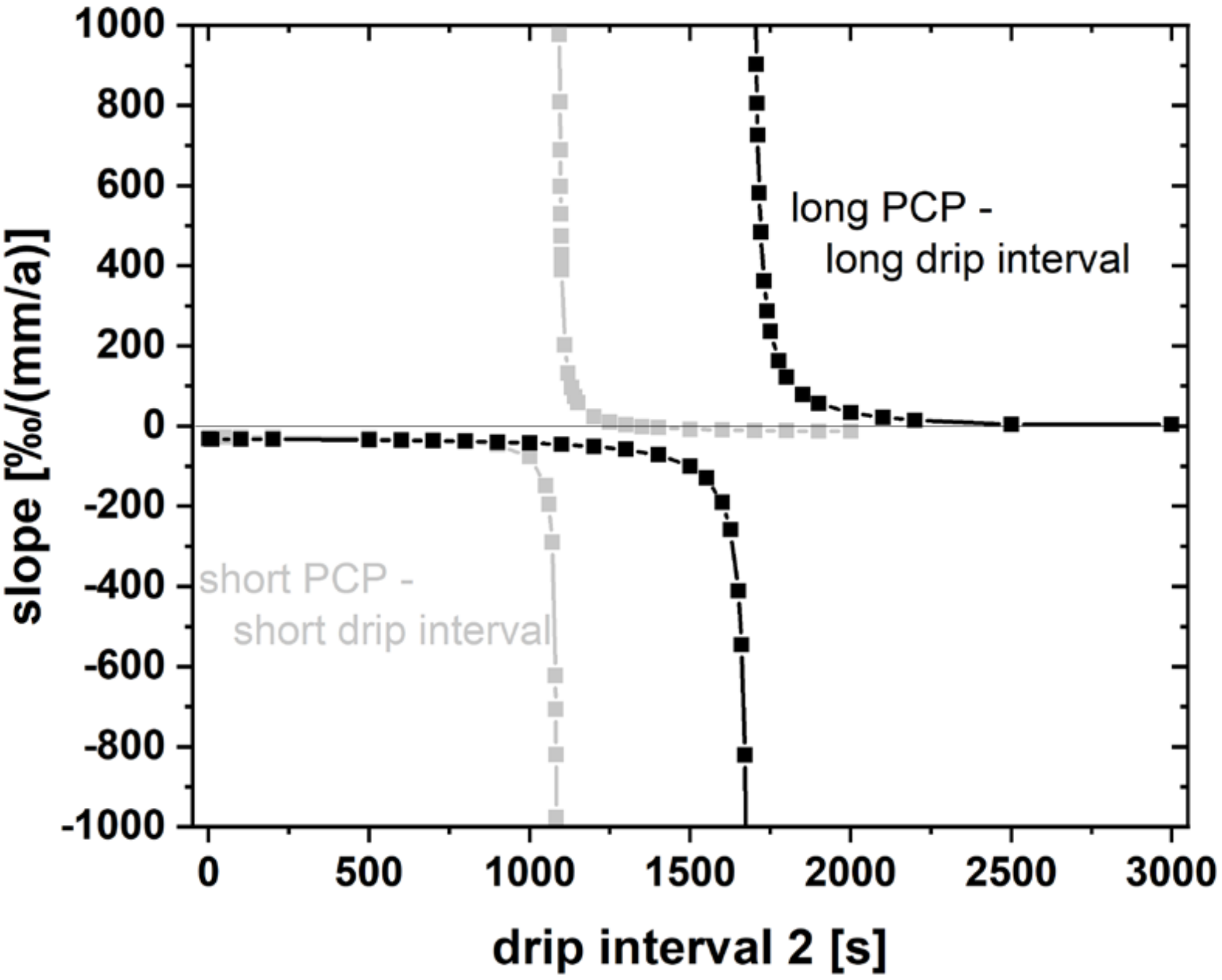

***Fig. 10:*** *The slope of* $\delta^{13}$C *over growth rate for two contemporaneously growing speleothems over drip interval variations of the second speleothem for two examples. Example 1 (grey): "short PCP – short drip interval" describes the situation where the time for PCP was fixed at drip site 1 to 100 s and at drip site 2 to 0 s, while drip interval 1 was set to 1s (drip interval 2 varies). Example 2 (black): "long PCP – long drip interval" describes the situation where the time for PCP is 1000 s for drip site 1 and 500s for drip site 2, while keeping the drip interval of site 1 at 500 s. Temperature is 5°C and cave air* $pCO_2$ *is 1000 ppm. Furthermore, initial drip water* $\delta^{13}$C composition *is -10 ‰ and* [$Ca^{2+}$] *is 2 moles/*$m^3$*, which corresponds to a solution in equilibrium with approximately 5 000 ppm of* $CO_2$*. The* $\delta^{13}$C *vs growth rate slope vary over a large range and changes the sign.*

When drip site 1 experiences a longer period of PCP and a longer drip interval than drip site 2, a complex hyperbola-like behaviour of the slope is observed (Fig. 10). The asymptotes, which are parallel to the x-axis, have a range in the slope as observed earlier in the modelling approach (Figs. 8, 9) and as most of the analysed pairs of speleothems. However, with longer drip intervals the slope rapidly decreases and reaches very large values, which are also found, albeit rarely, in our dataset, e.g., Abaco Island Cave, Bahamas (-2163 ‰/(mm/a); Arienzo et al., 2017) or Kesang Cave, China (-7462 ‰/(mm/a); Cheng et al., 2016). At even longer drip intervals, strongly positive slopes appear and rapidly decreases to either an asymptote with a slightly negative (-14 ‰/(mm/a); grey, Fig. 10) or slightly positive value for the slope (4 ‰/(mm/a); black, Fig. 10). Similar to those two examples, there are many further cases possible, where the same behaviour is observed. Similar as explained for the examples calculated without accounting for PCP or an equal PCP for the two drip sites, this behaviour is a result of the interplay between the rate of changes for $\delta^{13}C$ values and [$Ca^{2+}$] during $CO_2$ degassing and $CaCO_3$ precipitation.

This result provides a possible mechanism for the positive slopes observed in the SISAL extracted $\delta^{13}C$ data for contemporaneously growing speleothems from the same cave. In addition to this, our model experiments with different time periods of PCP and drip interval can also explain the large positive and negative slopes observed in some of the speleothem data (Fig. 3B). As the range of combinations of time periods for PCP and drip intervals, where such large positive or negative values are observed, is relatively small (only a few 100 seconds for drip interval 2), this might also provide an explanation why only few speleothem pairs show such large slopes.

From our modelling results we can now more confidently interpret the processes driving the slope between $\delta^{13}C$ data and growth rate. The slope is a measure of differences in the amount of fractionation between two contemporaneously growing speleothems, with which we can quantify the influence of cave processes on individual $\delta^{13}C$ time series. Analysing the slope in reproduced $\delta^{13}C$ time series of speleothems will give additional information on drip site characteristics. We found that the slope is sensitive to temperature, $pCO_2$, drip interval and the duration of PCP, but together with other proxies and the individual $\delta^{13}C$ time series, a more thorough evaluation of climatic conditions can be drawn.

5.5 Combination of $CaCO_3$ dissolution and re-precipitation

Another possibility to obtain positive slopes between $\delta^{13}C$ values and growth rates of two contemporaneously growing speleothems is through the combined effect of carbonate dissolution under different open-closed conditions and Rayleigh fractionation during PCP or CaCO3 precipitation on the stalagmite top. Let us assume that the host rock is dissolved in nearly completely closed carbonate conditions with only slightly different degrees of openness between two speleothem drip sites (e.g., gas volume = 1 and 2 l; Fig. 7). For a drip interval of 1 s for those two sites we obtained negative slopes (Sec. 5.3). Here, we want to investigate the effects on the slope, if drip intervals are considerably longer.

For different gas volumes of 1 and 2 l and a drip interval of 1 s for both stalagmite sites, the slope was about -6 ‰/(mm/a). When drip interval was about 100 s the slope increased to -5 ‰/(mm/a) and to +16 ‰/(mm/a) if one drip is falling each 2000 s. Thus, the combined effect of different carbonate dissolution conditions near the completely closed system and fractionation effects is also able to provide positive slopes.

When doing the same exercise for a nearly completely open carbonate dissolution system (gas volume = 200 and 500 l; Fig. 7) for drip intervals of 1, 100 and 2000 s, the slopes for $\delta^{13}C$ vs growth rates are -63, -67 and -170 ‰/(mm/a). Thus, positive slopes cannot be obtained by the combined effect of dissolution and re-precipitation under nearly completely open dissolution conditions.

In a last case, we do the calculations for two speleothems, where the carbonate dissolution occurred during nearly completely open and completely closed conditions (gas volume = 1 and 500 l; Fig. 7). For drip intervals of 1, 100 and 2000 seconds the slope remains always negative (-19, -20, -36 ‰/(mm/a)). However, as those extreme dissolution conditions were never observed, we render this calculation only as a hypothetical example. Further, very different carbonate dissolution conditions have not been observed within the same cave. In addition, in most caves carbonate dissolution conditions occur mostly under more open conditions (e.g., Genty et al., 1998; Griffiths et al., 2012; Lechleitner et al. ,2016), where it is not possible to obtain positive slopes for $\delta^{13}C$ values and growth rates, when combined with Rayleigh fractionation effects. Only in the rare cases where dissolution occurs under near-completely closed conditions, the combined effects of dissolution and Rayleigh effects can produce positive slopes. Radiocarbon measurements on each of the two stalagmites can reveal the carbonate dissolution conditions and thus can help to guide through the choice of which effect is responsible for a positive slope (dissolution and Rayleigh fractionation or the effect of decoupled length of PCP and drip intervals).

## 6. Conclusions

We have discussed the main factors influencing speleothem $\delta^{13}C$ values. With the $\delta^{13}C$ records extracted from the SISAL_v1 database and modelling results we were able to disentangle and quantify various processes affecting speleothem $\delta^{13}C$ values from a large number of speleothems from globally distributed caves. First, we focused on average $\delta^{13}C$ values of recently grown speleothems, accounting only for material deposited post-1900 CE. We found that $\delta^{13}C$ values are mainly affected by vegetation cover and temperature, but both relationships are subject to noise introduced by competing additional effects. The $\delta^{13}C$ - temperature relationship can be explained by temperature-vegetation and temperature-soil respiration rate dependencies, with additional modulation by the amount of precipitation for monsoonal areas. Although we applied a comprehensive approach, we nevertheless found extreme cases, which deviate from the main relationships. This is especially conspicuous for caves which are shown to have deep carbon sources or extreme ventilation that drives fractionation processes.

In the second part of our analysis, we focused on contemporaneously growing speleothems from individual caves. We observed a bimodal distribution of positive and negative slopes of $\delta^{13}C$ values vs growth rates. While negative slopes are expected, based on previously published in-cave fractionation model studies, positive slopes were more difficult to explain. By considering $CO_2$ degassing and $CaCO_3$ precipitation effects also for PCP and accounting for drip interval, we extended the Rayleigh fractionation approach of earlier in-cave fractionation models and we demonstrated that positive slopes between $\delta^{13}C$ values and growth rate can be explained by decoupling the time available for PCP and the drip interval for the individual drip sites or by the combined effect of nearly closed carbonate dissolution systems and long drip intervals.

Our data-model intercomparison highlights the various influences of in-situ processes and external climate conditions on speleothem $\delta^{13}C$ values. Not surprisingly, vegetation cover is an important driver of speleothem $\delta^{13}C$ values, but temperature emerges as a second factor that has a large effect, likely via its influence on the soil respiration rate. Soil respiration rate however, is also affected by the amount of precipitation, which was shown to be an important factor in warm regions. Furthermore, we showed that fractionation effects, especially via PCP, is important to explain the results of $\delta^{13}C$ differences in contemporaneously growing speleothems from the same cave. We propose that PCP should be implemented in next-generation CaCO3 precipitation models, as this process can explain much of the variations observed in our analysed dataset. Also, for climate reconstruction this process should be discussed in more detail, especially in combination with Mg/Ca or Sr/Ca ratios - $\delta^{13}C$ variations are a powerful tool to evaluate the strength of this process.

**Data availability**

The used speleothem data were extracted from the SISAL_v1 data base obtained from their repository (http://dx.doi.org/10.17864/1947.147). The code for extraction can be downloaded from the supplemental material. The extended Rayleigh-model for in-cave fractionation models are provided as an excel file in the supplement to this contribution or on request to the corresponding author. Eight free parameters can be used to fit the model to the cave conditions - temperature, initial and equilibrium [$Ca^{2+}$] as well as initial $\delta^{13}C$ composition. In addition, the time for PCP and drip interval for the two drip sites can be adjusted.

**Acknowledgements**

The authors acknowledge the speleothem community effort to establish the SISAL database, without which this contribution would not be possible. SISAL is an international working group of the Past Global Changes (PAGES) programme, and the authors gratefully acknowledge their support of this activity. JF is supported by DFG grant FO 809/4-1. NRGV is currently supported by EU-HORIZON Marie Curie Fellowship no. 796707. FAL gratefully acknowledges support from the Swiss National Science Foundation (SNSF) grant P400P2_180789. This project is TiPES contribution 9: This project has received funding from the European Union's Horizon 2020 research and innovation programme under grant agreement No 820970.